\documentclass{article}

\usepackage{arxiv}

\usepackage[utf8]{inputenc} 
\usepackage[T1]{fontenc}    
\usepackage{hyperref}       
\usepackage{url}            
\usepackage{booktabs}       
\usepackage{amsfonts}       
\usepackage{nicefrac}       
\usepackage{microtype}      
\usepackage{lipsum}		
\usepackage{graphicx}
\usepackage{natbib}
\usepackage{doi}

\usepackage{subfig}
\usepackage{amsmath}
\usepackage{siunitx}
\usepackage{booktabs}
\usepackage{multirow}
\DeclareSIUnit{\ohm}{\text{$\Omega$}}

\title{Dielectric Elastomer Actuator biased by Magnetorheological Elastomer with Permanent Magnet}

\author{ \href{https://orcid.org/0000-0001-8890-7157}{\includegraphics[scale=0.06]{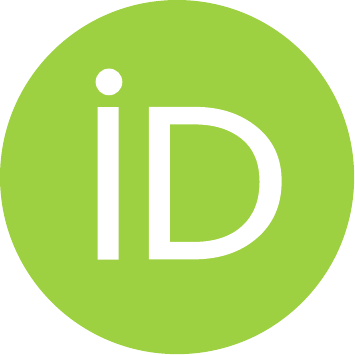}\hspace{1mm}Jakub Bernat} \\
	Institute of Automatic Control and Robotics\\
	Poznan University of Technology\\
	Piotrowo 3a, 60-965 Poznan, Poland \\
	\texttt{jakub.bernat@put.poznan.pl} \\
	\And
	\href{https://orcid.org/0000-0002-2177-1555}{\includegraphics[scale=0.06]{orcid.pdf}\hspace{1mm}Jakub Kołota} \\
	Institute of Automatic Control and Robotics\\
	Poznan University of Technology\\
	Piotrowo 3a, 60-965 Poznan, Poland \\
	\texttt{jakub.kolota@put.poznan.pl} \\
	\And
	\href{https://orcid.org/0000-0002-0229-3922}{\includegraphics[scale=0.06]{orcid.pdf}\hspace{1mm}Piotr Gajewski} \\
	Institute of Chemical Technology and Engineering\\
        Poznan University of Technology\\
	Berdychowo 4, 60-965 Poznan, Poland \\
	\texttt{piotr.gajewski@put.poznan.pl} \\
        \And
	\href{https://orcid.org/0000-0002-8299-5085}{\includegraphics[scale=0.06]{orcid.pdf}\hspace{1mm}Agnieszka Marcinkowska} \\
	Institute of Chemical Technology and Engineering\\
	Poznan University of Technology\\
	Berdychowo 4, 60-965 Poznan, Poland \\
	\texttt{agnieszka.marcinkowska@put.poznan.pl}
}

\renewcommand{\shorttitle}{DE Actuator biased by MRE with Permanent Magnet}

\hypersetup{
pdftitle={A template for the arxiv style},
pdfsubject={q-bio.NC, q-bio.QM},
pdfauthor={David S.~Hippocampus, Elias D.~Striatum},
pdfkeywords={First keyword, Second keyword, More},
}

\begin{document}
\maketitle

\begin{abstract}
Dielectric elastomer actuators have become one of the most important smart material transducers in recent times. One of the crucial aspects in this field is the application of bias to find the best operating conditions. The basic task is to find the proper bias configuration to obtain a wide range of displacements in the actuator. In the literature, various biases, such as mechanical springs, permanent magnets, or pneumatic springs, are studied. In our work, the magnetorheological elastomer is applied to build a novel bias that ensures a wide range of displacement. Because of the softness and the compliant chemical structure, the magnetorheological elastomer can be easily integrated with the dielectric elastomer actuator. The magnetorheological elastomer as a bias for a dielectric elastomer actuator is verified in the series of experiments. Finally, the discussion on the advantages and disadvantages of the new bias type is performed.
\end{abstract}

\keywords{Dielectric Elastomer Actuator (DEA) \and Magnetorheological Elastomers (MRE) \and Bias \and Dielectric Electroactive Polymers (DEAP)}

\section{Introduction}
The dynamic development of soft materials has recently allowed for the construction of new, interesting actuators, sensors, and generators \cite{Rus:2015,Chen:2020:softrobots}. One of such solutions are DEAP (Dielectric Electroactive Polymers) membrane, which belong to the group of smart materials. The actuators made of DEAP are characterized by a very flexible structure, high value of strain, and fast actuation response time \cite{Rosset:2016,Bernat:applied:13031651}. The recent practical developments like pumps for soft robotics, speakers, grippers, and haptic displays show the wide range of applicability of these devices \cite{Cao2019,Anderson2012,Zhang2015,Hau:201824,Phung:2021}.

DEAP actuators are composite materials that consist of an elastomer film covered by electrode layers on both sides. In order to define the operating range of such actuators, a bias is used in addition to the initial stretching of the actuator membrane. The biasing mechanism is used to pre-load the membrane, allowing stroke in the out-of-plane direction. In the literature, a few concepts of bias are known: gravitational, linear spring, nonlinear mechanical spring, and permanent magnets. The first uses mass to perform a load on the membrane. However, the biasing mass increases the overall size and weight of the actuator, and the principle of utilizing gravity restricts its application (the requirement to keep the device horizontal) \cite{Sarban:5783345,Bernat:IEEE:2021}. The linear spring allows for independence from gravity but does not provide large deformations \cite{Rizzello:6867294,Rizzello:Nlr:2015}. The large displacement can be obtained for nonlinear mechanical springs where the linear spring and nonlinear spring are coupled \cite{Rizzello:Nlr:2015}. Another non-contact bias system that provides large displacement is two permanent magnets \cite{Kolota:2020,PhysRevApplied.12.044033}. In literature \cite{Rizzello_2015, 10.1115/1.4028456,    SASIKALA201758}, approximated mathematical models have been developed to characterize the performance of conical DEAP actuators with biasing springs and biasing mass. 

In this work, a new concept combining  Permanent Magnet (PM) and Magnetorheological Elastomers (MRE) bias for DEAP actuator is described where a soft MRE disc is integrated with DEAP membrane. The deflection mechanism is achieved by magnetic force from a permanent magnet, which pulls the central MRE disk out of the plane, creating a conical geometry of the actuator. The innovative solution is characterized by the preservation of the flexible structure of the actuator and a significant extension of the working range resulting from the non-linear nature of the force coming from the permanent magnet. 

The paper is organized as follows. Descriptions of the PM-MRE bias model, its operating principle, and its benefits are given in Section 2. Section 3 of the paper presents the Finite Element Model (FEM) model of bias structure and the results of the simulation of the force between the MRE disc and the permanent magnet for a variable gap. A description of the MRE and DEAP preparation process as well as key material properties are presented in Section 4. Extensive experiments results for transients and steady-state characteristics are presented in Section 5. There are included the comparison with classical mass bias and the detailed analysis for several series with different values of the offset parameter. The paper ends with concluding remarks in Section 6.

\section{Innovative concept of flexible actuator bias}

\begin{figure}
  \centering
   \includegraphics[width=0.4\textwidth]{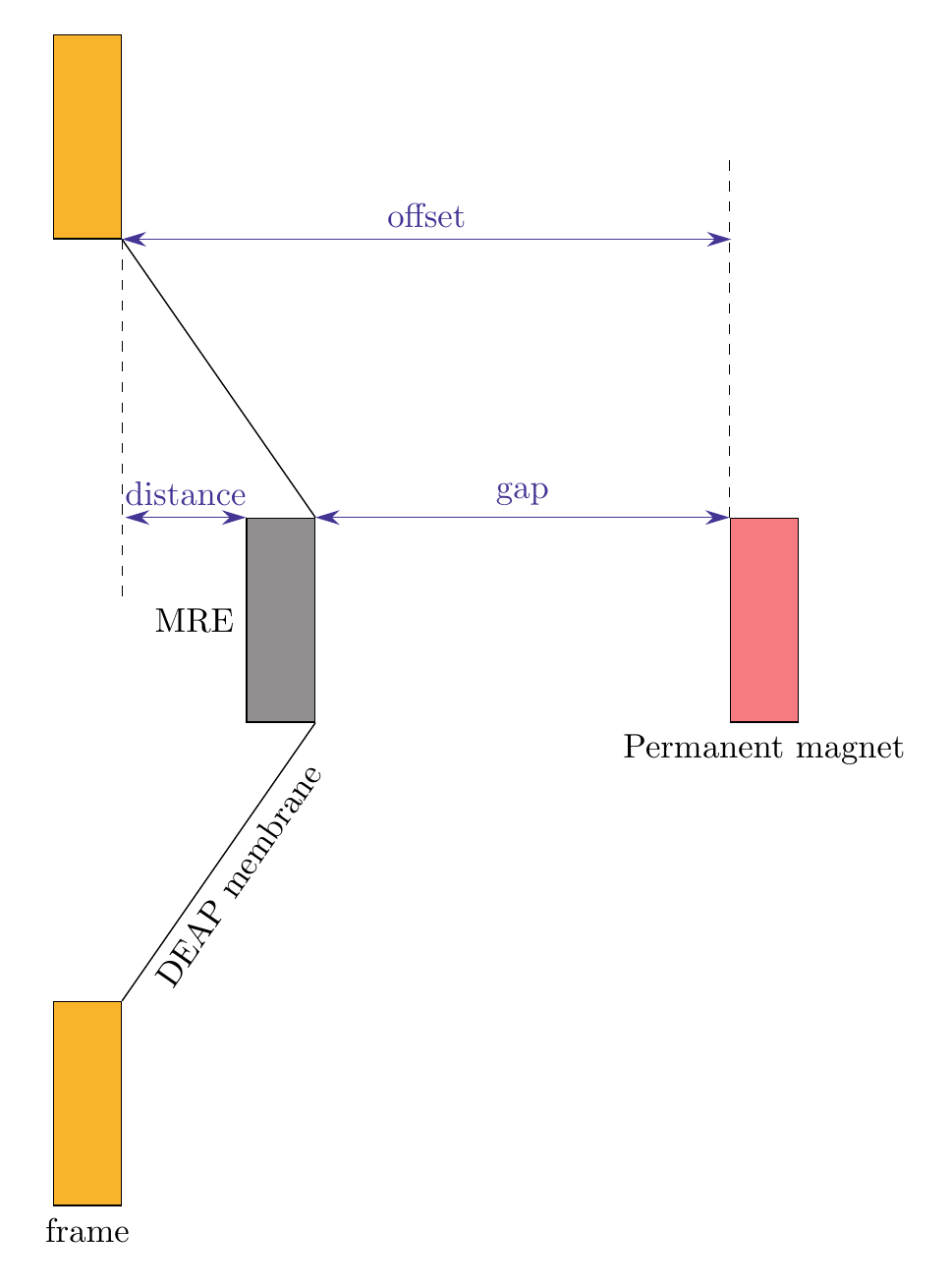}
  \caption{The DEAP actuator with PM-MRE configuration.}
  \label{fig:actuator}
\end{figure}

In our work, we introduce the DEAP actuator biased by force generated between a permanent magnet and a magnetorheological elastomer. The construction of the actuator is shown in Figure \ref{fig:actuator} and consists of a fixed frame, DEAP membrane with mounted MRE disc, and permanent magnet. The magnetic field of the permanent magnet causes MRE to be pulled to the magnet. In normal operation mode, if the voltage is turned on, the membrane is less stiff, and hence MRE is moved closer to MRE. If the voltage is turned off, the membrane restores to the initial position. The exact range of displacement depends on the force versus displacement characteristics for DEAP, MRE, and permanent magnet.

\begin{figure*}[!t]
\centering
\subfloat[]{\includegraphics[width=0.45\textwidth]{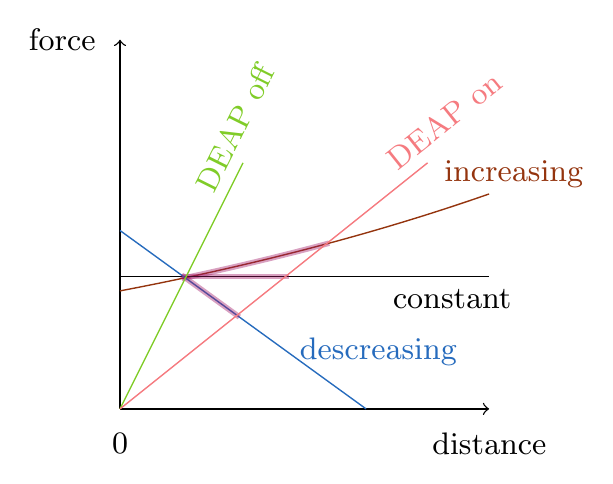}
}
\hfil
\subfloat[]{\includegraphics[width=0.45\textwidth]{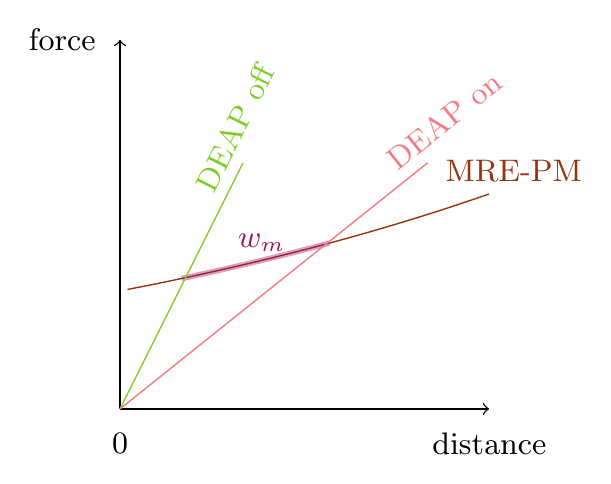}
}
\caption{The concept of decreasing, constant and increasing force bias (a). 
The concept of MRE-PM bias. The working range for MRE-PM system is denoted as $w_m$ (b). }
\label{fig:biases}
\end{figure*}

The concept of bias is to make two different equilibria when DEAP is turned on and turned off. Therefore, it is important to understand the characteristics between forces acting on DEAP vs. displacement and bias force vs. displacement. The same problem arises for the biases described in the introduction; however, different types of forces are required to compare depending on bias components (i.e. linear spring, nonlinear spring, permanent magnets, or mass). Authors noticed that biases can be grouped into three types: decreasing, constant, and increasing. In Figure \ref{fig:biases}a the general force-displacement characteristics are shown with denoted working range. It is clear that the crucial aspect is to create an increasing pull force when a DEAP actuator is powered. Currently, such a possibility is for nonlinear mechanical bias and permanent magnet bias. 

We propose a new type of bias made by the connection of MRE-PM. The MRE material is soft, and it enables easy integration with DEAP. Furthermore, the force between MRE-PM also has increasing characteristics that allow one to extend its working range, as shown in Figure \ref{fig:biases}b and as will be shown in the next part of the work.

\begin{table}
 \caption{The summary of bias type in literature.}
  \centering
  \begin{tabular}{ccc}
    \toprule
    Name     & Example Works & Characteristics Type \\
    \midrule
    linear spring & \cite{Rizzello:6867294,Rizzello:Nlr:2015} & decreasing  \\
    pneumatic & \cite{Bernat:Pneumatic:2021} & decreasing \\
    gravitational & \cite{Sarban:5783345,Bernat:IEEE:2021} & constant \\
    PM-PM & \cite{Kolota:2020,PhysRevApplied.12.044033} & increasing \\
    nonlinear spring (bi-stable) & \cite{Rizzello:Nlr:2015} & increasing \\
    \bottomrule
  \end{tabular}
  \label{tab:table}
\end{table}

\section{FEM model of coupling MRE-PM bias system}

In this part, the characteristics of force between the permanent magnet and magnetorheological elastomer disc are analyzed. As stated above, the shape of the characteristics is crucial from the point of view of bias.

\begin{figure}
  \centering
   \includegraphics[width=0.7\textwidth]{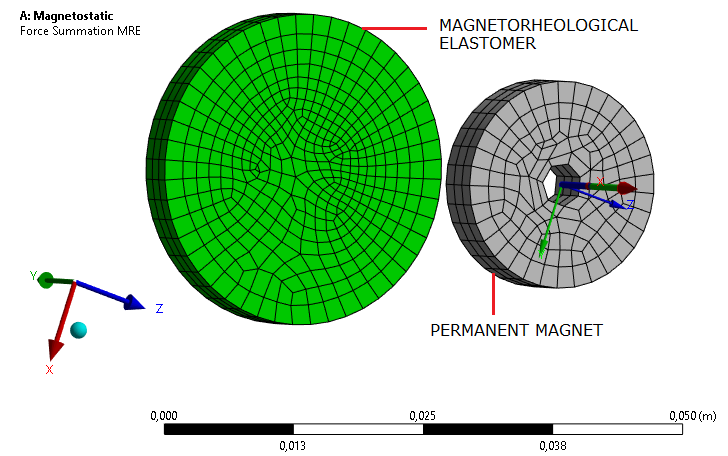}
  \caption{The configuration of ANSYS FEM simulation of the force between the MRE disc and the permanent magnet for a variable gap.}
  \label{fig:fem:definition}
\end{figure}

\begin{figure}
  \centering
   \includegraphics[width=0.7\textwidth]{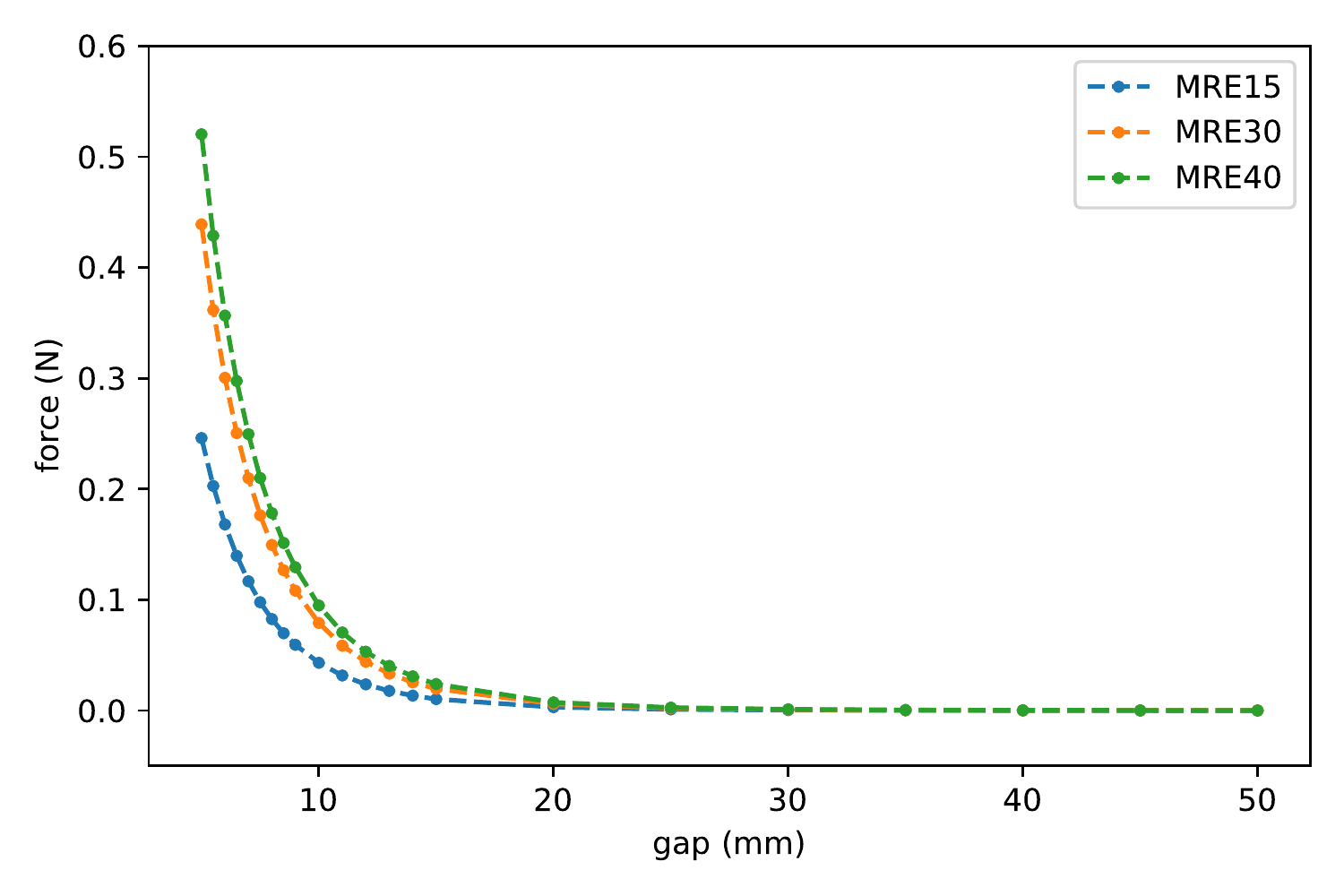}
  \caption{The results of FEM simulation of the force between the MRE disc and the permanent magnet for a variable gap.}
  \label{fig:fem:simulation}
\end{figure}

The analyzed MRE coupling system with a permanent magnet was modeled in the FEM simulation environment in an iterative manner with variable parameters using the magnetostatic module. The software package Ansys, Inc. Products 2022 R2 can be used for the investigation of the magnetic field distribution and basic electromagnetic characteristics. The finite elements method assures sufficient accuracy of electromagnetic field computation and very good flexibility when the geometry is modeled and field sources are loaded. A simulation was performed in order to obtain the force characteristics between the MRE disc and the permanent magnet for a variable gap. The gap and the calculated force were determined in the center of both tested elements. The simulation was carried out for 22 measurement points in the range of changes in the position of the magnet to the MRE disk corresponding to the experiments (from 50 mm to 5 mm). Figure \ref{fig:fem:definition} shows the 3D MRE-permanent magnet system with a quadratic computational mesh (42519 nodes and 25852 elements). The magnet was modeled with the dimensions and parameters presented in Table \ref{tab:magnetParameters}, which are consistent with the real object used in the experiment. The simulations were repeated for three different isotropic relative permeability (see Table \ref{tab:coil}) and the results are presented in Figure \ref{fig:fem:simulation}. The results show that MRE disc with a permanent magnet has nonlinear characteristics which are applicable to use in DEAP actuators. Furthermore, it is possible to set up DEAP membrane, MRE disc, and permanent magnet that the bias is increasing.

From our point of view, it is worth comparing the presented results with the two permanent magnets used as the bias known in the previous works \cite{PhysRevApplied.12.044033}. The characteristics obtained for two permanent magnets \cite{Liu:2021} and permanent magnet and magnetorheological elastomer (obtained in this work) have similar shapes. However, it is worth pointing out that two magnets have much more accumulated energy than permanent magnet-MRE with the same shape, hence the same force can be obtained with smaller PMs. On the other hand, the permanent magnet is a rigid element while magnetorheological elastomer is soft which is the main author's motivation to use in DEAP actuator.

\begin{table}
 \caption{The parameters of neodymium permanent magnet N38 used in bias.}
  \centering
  \begin{tabular}{lll}
    \toprule
    \multicolumn{2}{c}{Part}                   \\
    \cmidrule(r){1-2}
    Name     & Description     & Size \\
    \midrule
    outer diameter   & $r_{1}$  & \SI{20.0}{\milli\meter}     \\
    inner diameter   & $r_{2}$  & \SI{4.2}{\milli\meter}      \\
    thickness        & $w$      & \SI{5.0} {\milli\meter}  \\
    weight             & $m$        & \SI{11.26}{\gram}  \\
    density   & $d$  & \SI{7.5}{\gram/\centi\meter^{3}}      \\
    Vickers hardness        & $HV$      & \SI{600,0} {\kilo\gram/\milli\meter^{2}}  \\
    resistivity             & $R$        & \SI{144.0}{\micro\ohm \centi\meter}  \\
    coercive force            & $F_{c}$        & \SI{1.45E+05}{Am^{-1}}  \\
    residual induction            & $B_{r}$        & \SI{1.23}{T}  \\
    \bottomrule
  \end{tabular}
  \label{tab:magnetParameters}
\end{table}

\section{Fabrication process of MRE and DEAP}
\label{se:fabrication}
\textit{Preparation of MRE}

\begin{figure}
  \centering
   \includegraphics[width=0.4\textwidth]{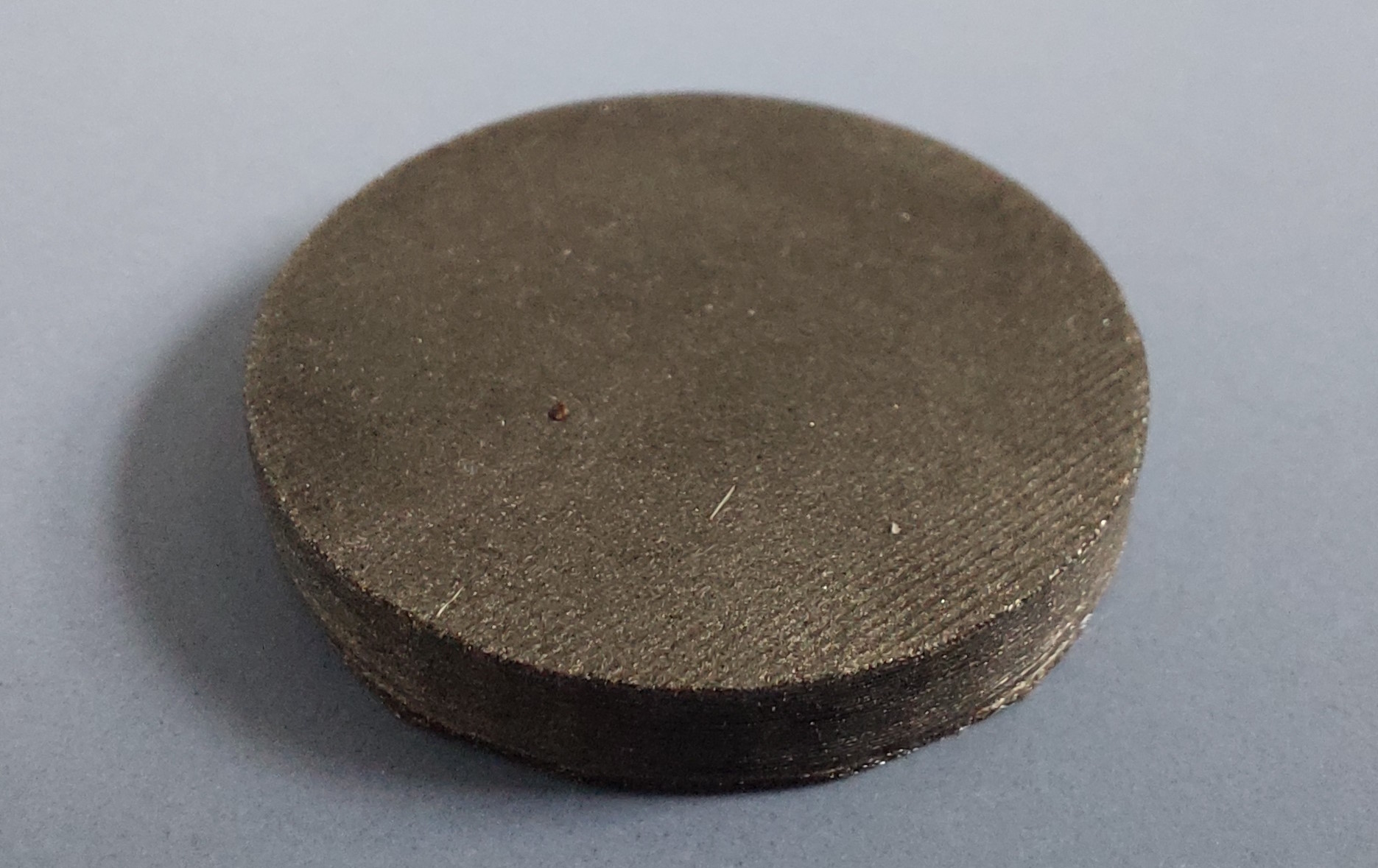}
  \caption{Sample of MRE disc.}
  \label{fig:MRE}
\end{figure}

The MRE materials were obtained according to the previously described procedure \cite{Bernat:sensors:22072757}. In the first step, silicone RTV-2 (OTT-S825 from OTTSilicone) and metal flake powder (FeSiAl, SP-3B from Mate Co., Ltd., Japan) with  a particle size of D50=35-55 \SI{}{\micro\meter} were weighed into a glass vessel. Next, components were mixed thoroughly until a homogeneous mixture was obtained, and the catalyst was added (2\% by weight). After vigorous mixing, the resulting mixture was degassed using a vacuum. In the second step, the prepared mixture was poured into a disc-shaped mold (diameter 20~mm, thickness 4~mm) and a beam-shaped mold (length 60~mm, width 15~mm, thickness 2~mm). The molds were made by 3D printing. After 24h MRE samples were removed from the molds and carefully checked for bubbles in the structure. Metal flake powder was used in various amounts, i.e. 15\% by weight (MRE15), 30\% by weight (MRE30), and 40\% by weight (MRE40). An example MRE disc is shown in Figure \ref{fig:MRE}.

\begin{table}
 \caption{The parameters of DEAP actuator.}
  \centering
  \begin{tabular}{lll}
    \toprule
    \multicolumn{2}{c}{Part}                   \\
    \cmidrule(r){1-2}
    Name     & Description     & Size \\
    \midrule
    Pre-stretch radius   & $R_{pre}$  & \SI{6}{\centi\meter}     \\
    Post-stretch radius  & $R_{post}$ & \SI{12}{\centi\meter}      \\
    Resistance            & $R$        & \SI{0.698}{\mega\ohm}  \\
    Capacity              & $C$        & \SI{2.2}{\nano\farad}  \\
    \bottomrule
  \end{tabular}
  \label{tab:actuatorParameters}
\end{table}

The properties of the obtained materials, i.e. Young's modulus $(E_{mod})$, and relative permeability were characterized. The Young modulus, as well as the relative permeability (two independent coils) of MRE materials, were measured with the procedure described in the work \cite{Bernat:sensors:22072757}. The results are presented in Table \ref{tab:coil}. As can be seen relative permeability as well as Young modulus increase with increasing amounts of metal flake powder. The Young modulus of silicone without the addition of metal flake powder has the value of \SI{0.31 \pm 0.03}{\mega\pascal}. Thus, the higher amount of the metal flake powder the higher stiffness of the MRE materials as visible in Table \ref{tab:mechanicalMRE}.
\begin{table}
 \caption{The relative permeability measured by two independent coils.}
  \centering
  \begin{tabular}{lcc}
    \toprule
    Material     & Coil A & Coil B \\
    \midrule
    MRE15   & \num{2.24}  & \num{2.18} \\
    MRE30   & \num{4.21}  & \num{4.12} \\
    MRE40   & \num{5.50}  & \num{5.32} \\
        \bottomrule
  \end{tabular}
  \label{tab:coil}
\end{table}
\begin{table}
 \caption{The Young modulus of prepared MRE materials.}
  \centering
  \begin{tabular}{cc}
    \toprule
     Material & Value \\
    \midrule
     MRE15 & \SI{0.85 \pm 0.05}{\mega\pascal} \\
     MRE30 & \SI{1.50 \pm 0.13}{\mega\pascal} \\
     MRE40 & \SI{2.20 \pm 0.20}{\mega\pascal} \\
    \bottomrule
  \end{tabular}
  \label{tab:mechanicalMRE}
\end{table}

\textit{Preparation of DEAP}

The DEAP actuator was made of 3M VHB tape as a radial membrane (radius \SI{60}{\milli\meter}) embedded in poly(methyl methacrylate) rings. The \SI{1}{\milli\meter} tape was pre-stretched from radius \SI{30}{\milli\meter} to \SI{60}{\milli\meter} and then carbon grease electrodes were applied. The measurement of the actuator LCR parameters gives a capacitance equal to \SI{2.2}{\nano\farad} and \SI{0.7}{\mega\ohm}. The Young modulus ($E_{mod}$), stress at 100\% strain ($\sigma_{100}$), stress ($\sigma_{max}$) and strain ($\epsilon_{max}$) at brake of VHB tape were measured. Obtained values are given in Table \ref{tab:mechanicalVHB}.
%
\begin{table}
 \caption{The mechanical parameters of VHB membrane.}
  \centering
  \begin{tabular}{ccc}
    \toprule
    Parameter & Symbol & Value \\
    \midrule
    Young Modulus & $E_{mod}$ & \SI{0.158 \pm 0.011}{\mega\pascal} \\
    stress at 100\% strain & $\sigma_{100}$ & \SI{0.095 \pm 0.003}{\mega\pascal} \\
    stress at brake & $\sigma_{max}$ & \SI{0.233 \pm 0.015}{\mega\pascal} \\
    strain at brake & $\epsilon_{max}$ & \SI{875 \pm 50}{\%} \\
    \bottomrule
  \end{tabular}
  \label{tab:mechanicalVHB}
\end{table}

\section{Experiments}

In this section, the experiments verifying the concept of actuators are performed. In the experimental analysis, three different MRE samples (i.e. MRE15, MRE30, and MRE40) are manufactured by varying the percentage of metal flake powder particles as described in Section \ref{se:fabrication}. First, the static characteristics of force between the MRE samples and the permanent magnet are obtained (without the usage of DEAP). Second, the static and dynamic operation of the actuator is measured.

\subsection{Static force-distance characteristics}

\begin{figure*}[!t]
\centering
\includegraphics[width=0.5\textwidth]{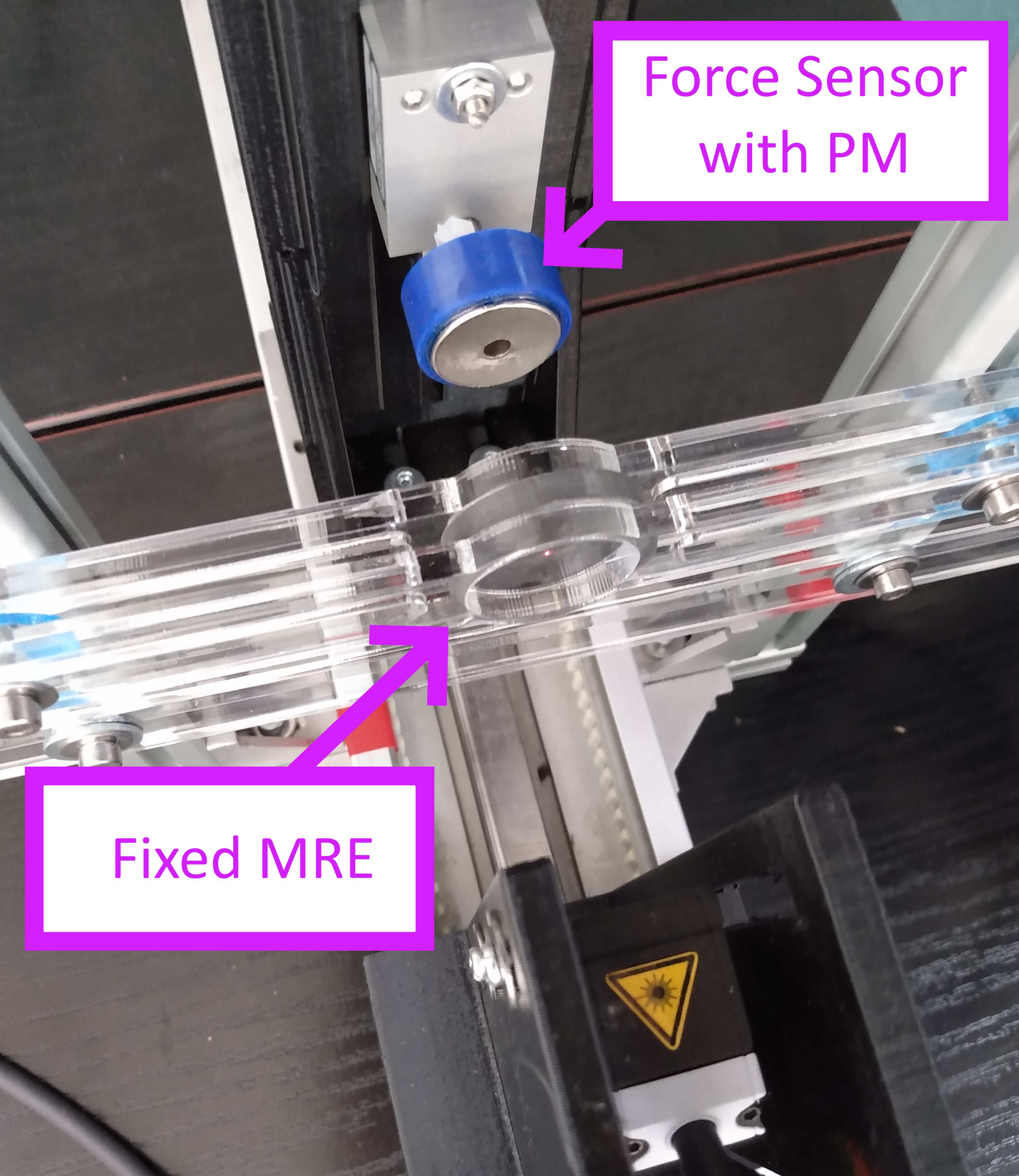}
\caption{The laboratory kit applied to measure characteristics between permanent magnet and MRE.}
\label{fig:mre-pm-kit}
\end{figure*}

\begin{figure*}[!t]
\centering
\subfloat[]{\includegraphics[width=0.45\textwidth]{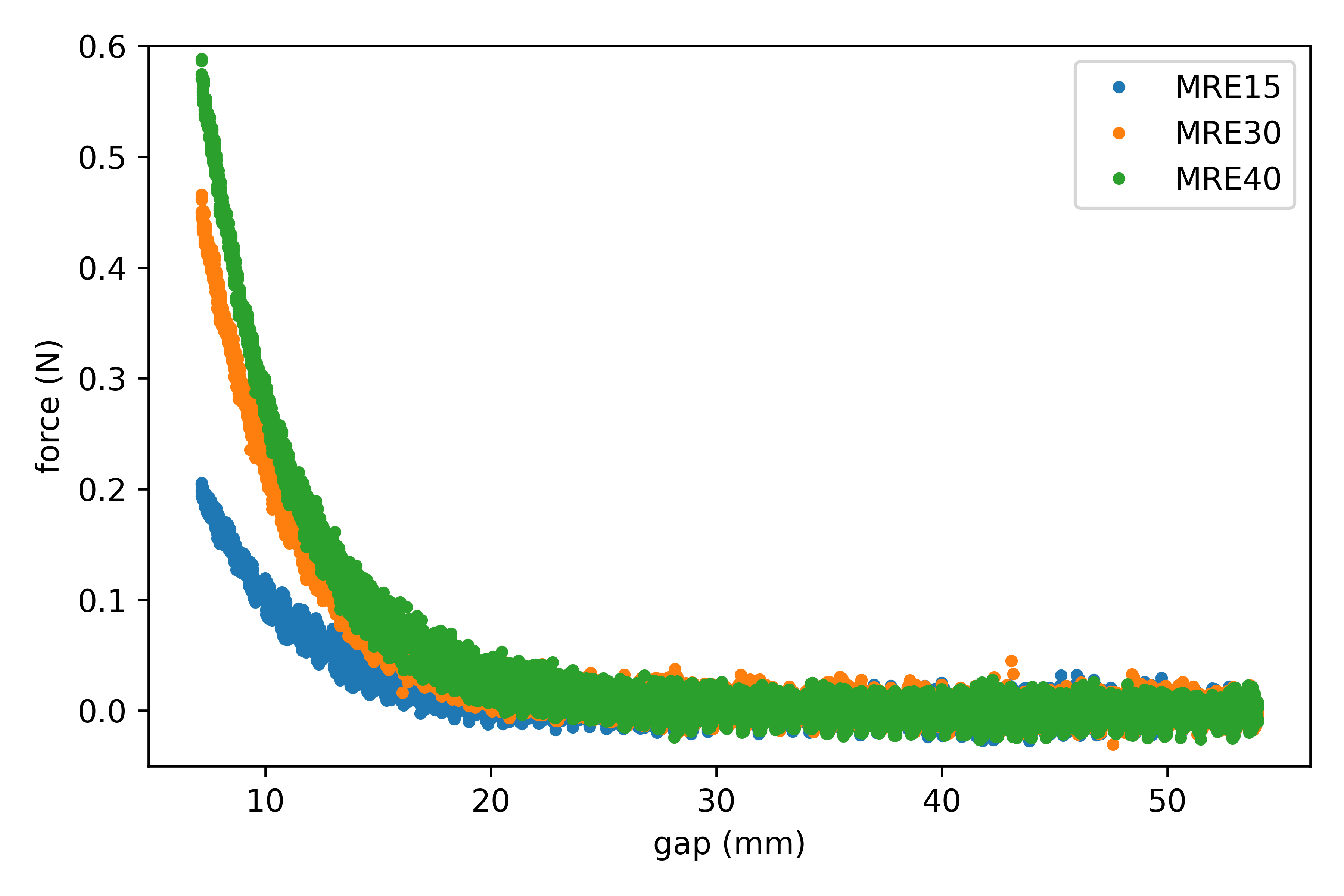}
}
\hfil
\subfloat[]{\includegraphics[width=0.45\textwidth]{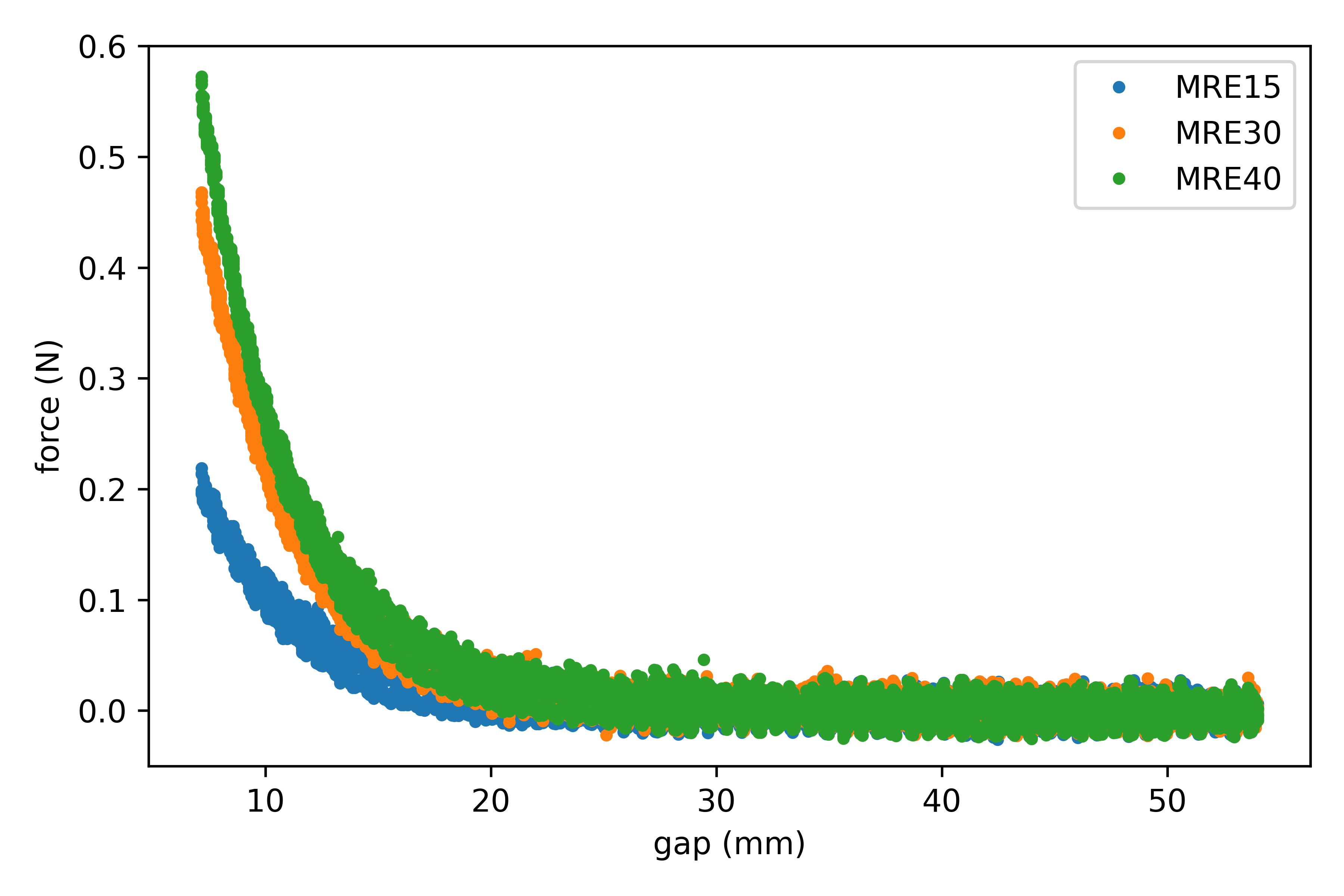}
}
\caption{The measurement of force between PM and MRE for various discs.}
\label{fig:forcePM}
\end{figure*}

In the first stage of experiments, the static force versus distance where obtained for MRE disc and permanent magnet. To measure the force characteristics between the magnetorheological elastomer and permanent magnet, three MRE discs are mounted on the fixed non-magnetic frame. Furthermore, the laboratory set shown in Figure \ref{fig:mre-pm-kit} included a precise linear module and a force sensor (Transducer Techniques GS-250) with a permanent magnet mounted on it. The force sensor measures the force acting between the permanent magnet and the magnetorheological disc. The gap between MRE and the permanent magnet is controlled by a linear module. Such an automated measurement environment made it possible to carry out two repeatable, identical series without human intervention. The signals measured by the platform were acquired to obtain force characteristics in the range of gap from \SI{50}{\milli\meter} to \SI{5}{\milli\meter}. Figure \ref{fig:forcePM}a  shows the results of the first measurement series, while Figure \ref{fig:forcePM}b presents the characteristics of the second measurement series for various MRE discs. The results obtained in the two series show good repeatability and prove the reproducibility of the experiments. The characteristics, as expected, show an exponential character depending on the gap between the magnet and MRE. In addition, it is getting stronger with increasing metal flake powder content in subsequent MRE discs. For MRE40, the peak force value achieved for a minimum distance of \SI{5}{\milli\meter} is \SI{0.6}{\newton}, while for MRE15 it is only 30 percent of this value and the force is \SI{0.2}{\newton}. The results obtained in the experiments are compatible with FEM simulations presented in Section 3. Furthermore, the experiments show that the characteristic is non-linear and can be exploited in the bias construction.

\subsection{The steady-state characteristics of DEAP actuator with PM-MRE bias}

\begin{figure*}[!t]
\centering
\includegraphics[width=0.5\textwidth]{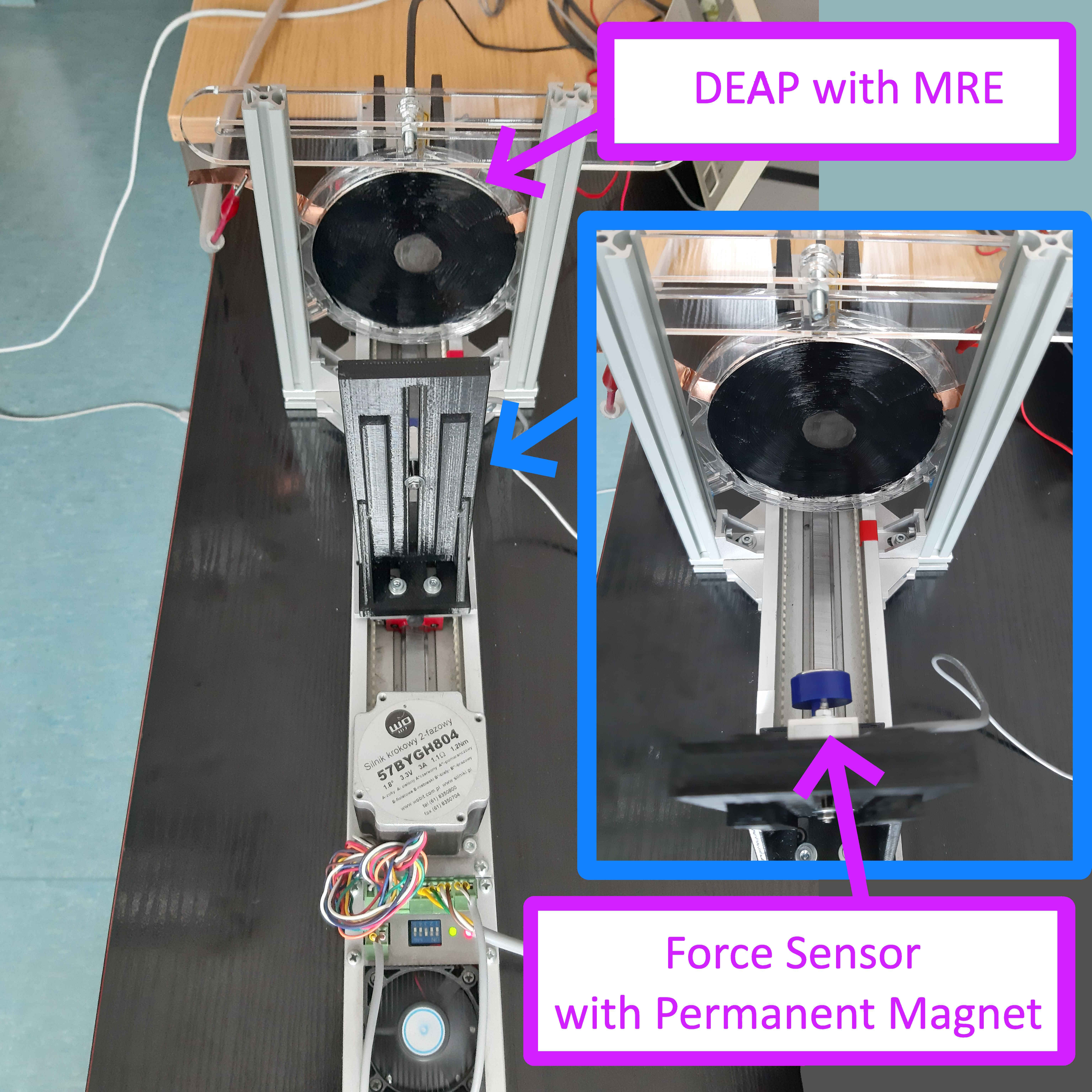}
\caption{The laboratory kit applied to DEAP actuator with MRE-PM bias.}
\label{fig:mre-deap-pm-kit}
\end{figure*}

\begin{figure*}[!t]
\centering
\subfloat[]{\includegraphics[width=0.45\textwidth]{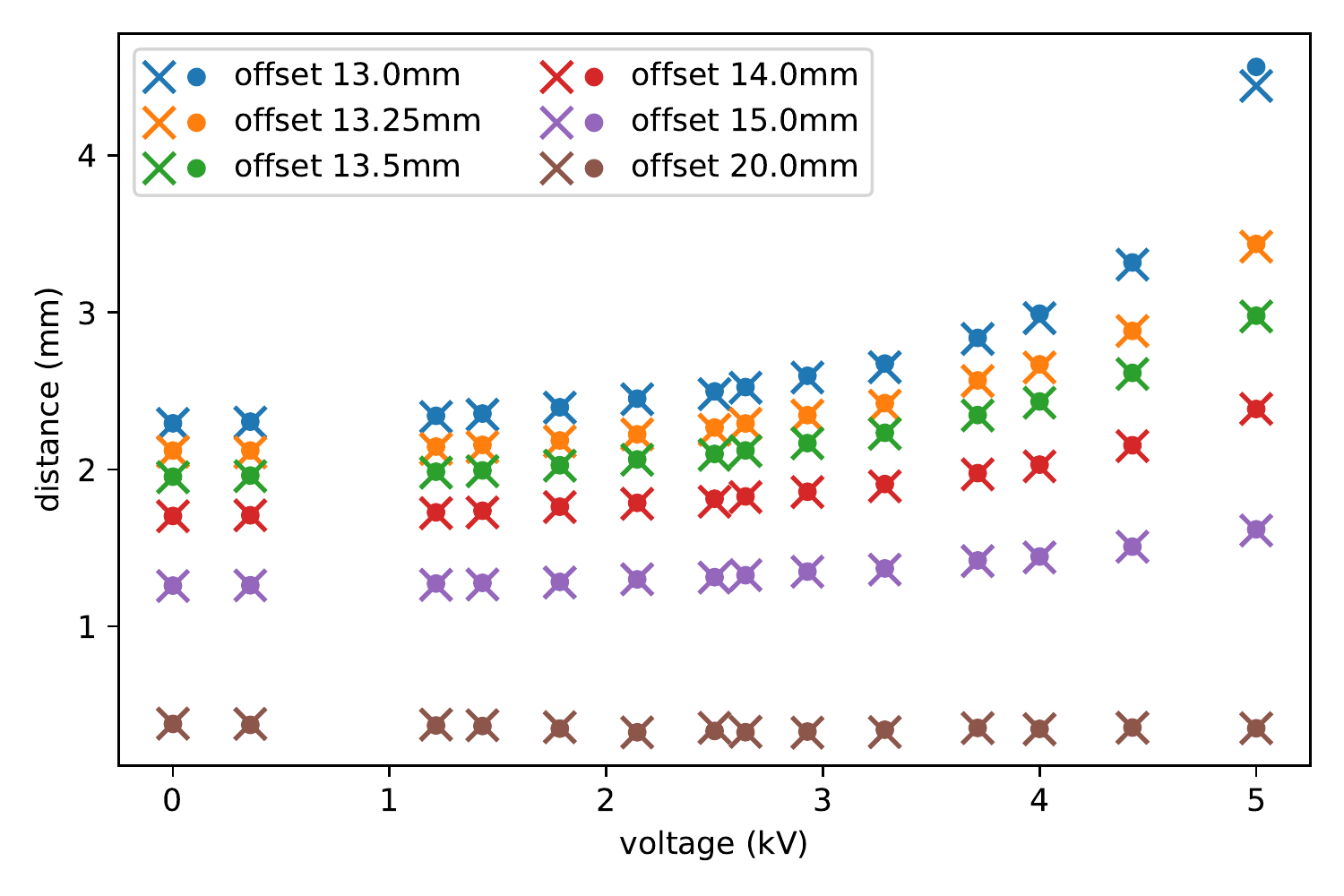}
}
\hfil
\subfloat[]{\includegraphics[width=0.45\textwidth]{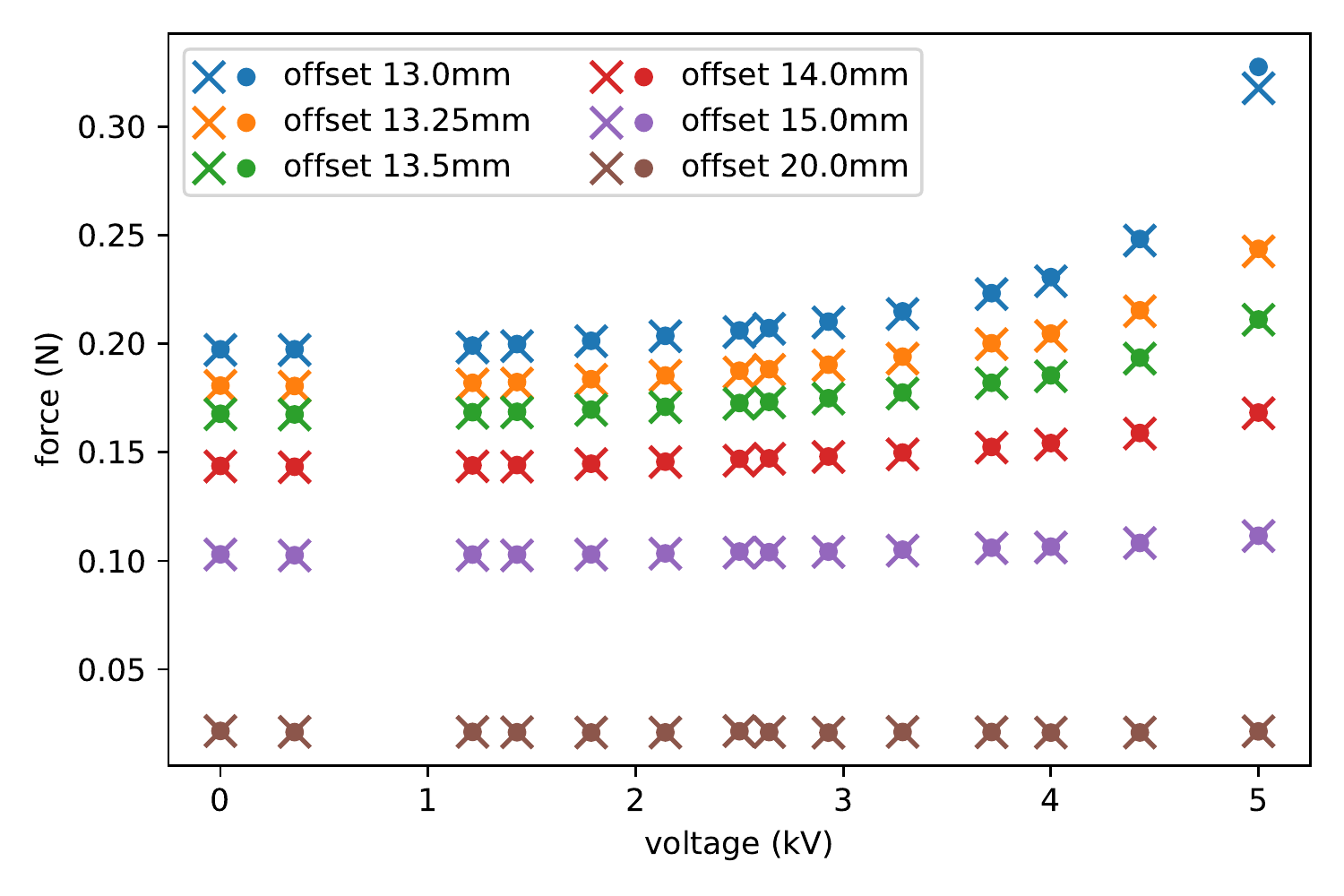}
}
\\
\subfloat[]{\includegraphics[width=0.45\textwidth]{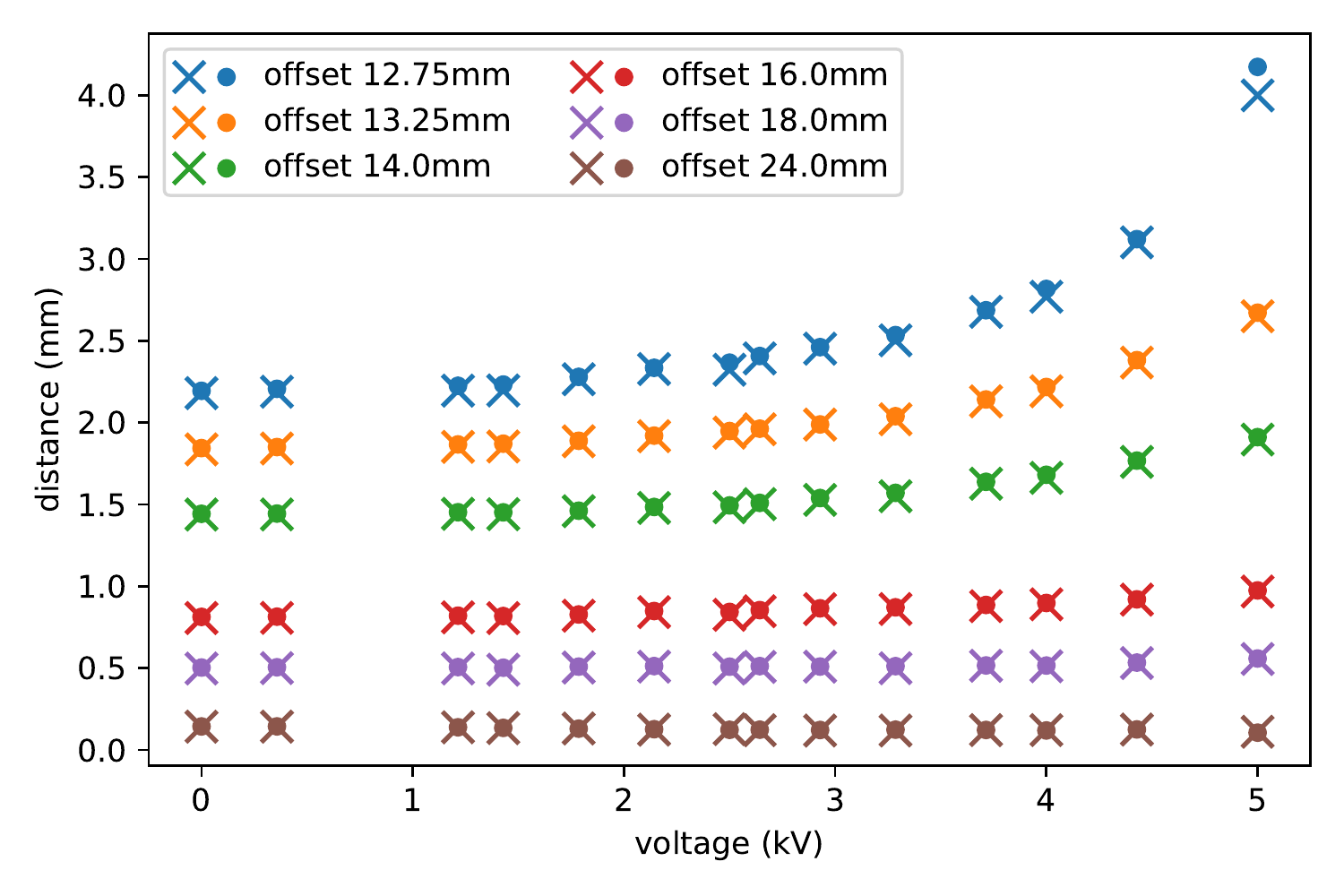}
}
\hfil
\subfloat[]{\includegraphics[width=0.45\textwidth]{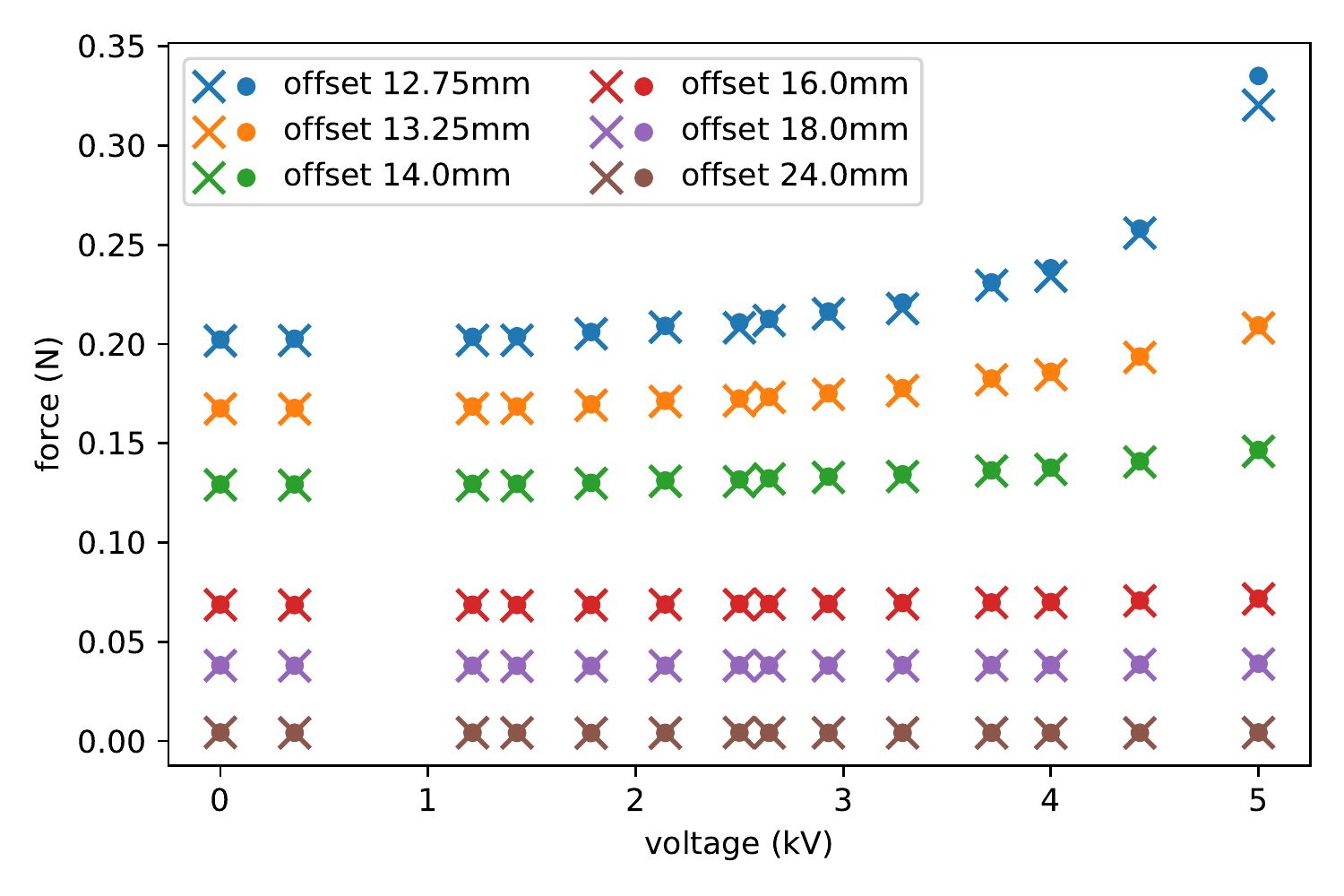}
}
\\
\subfloat[]{\includegraphics[width=0.45\textwidth]{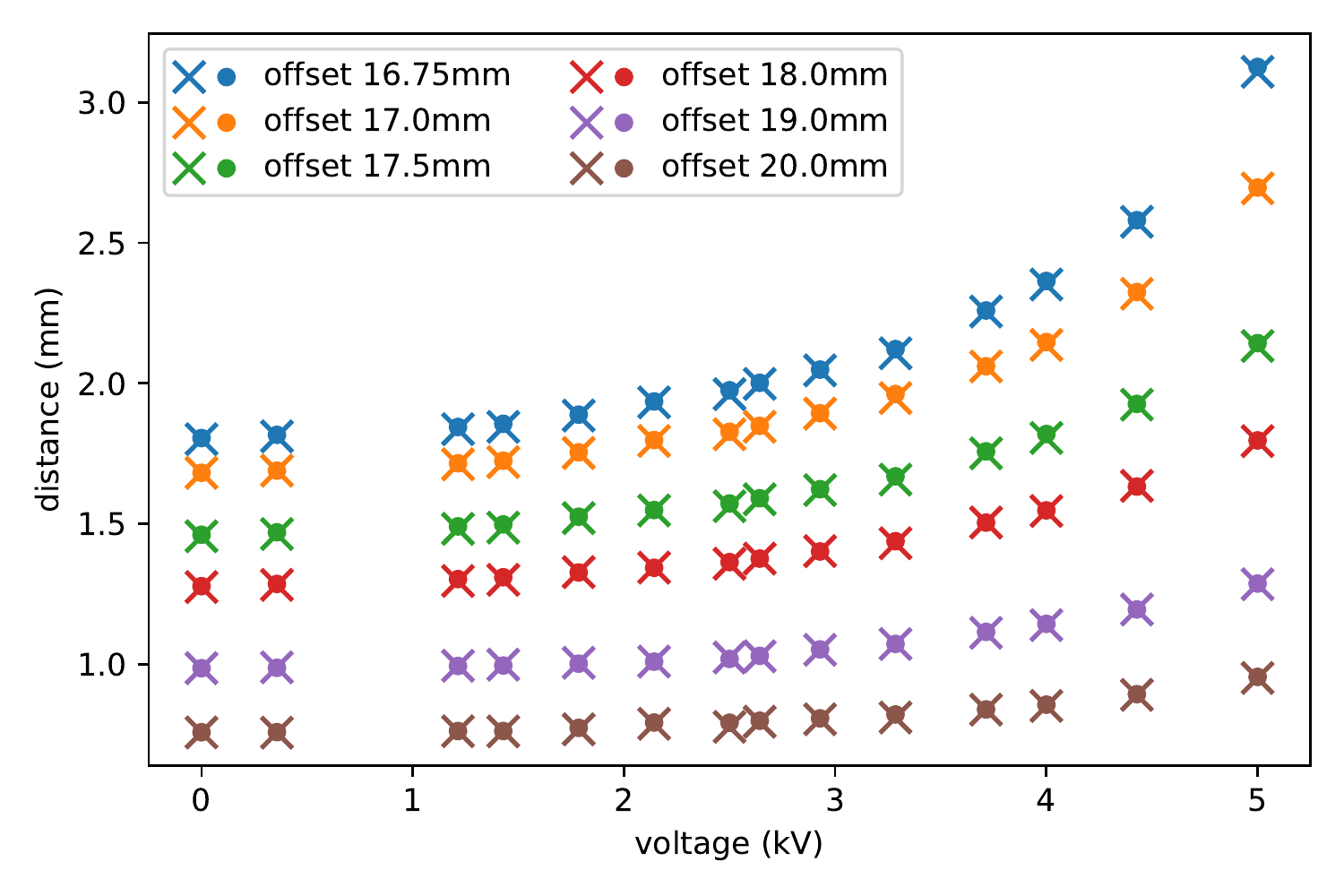}
}
\hfil
\subfloat[]{\includegraphics[width=0.45\textwidth]{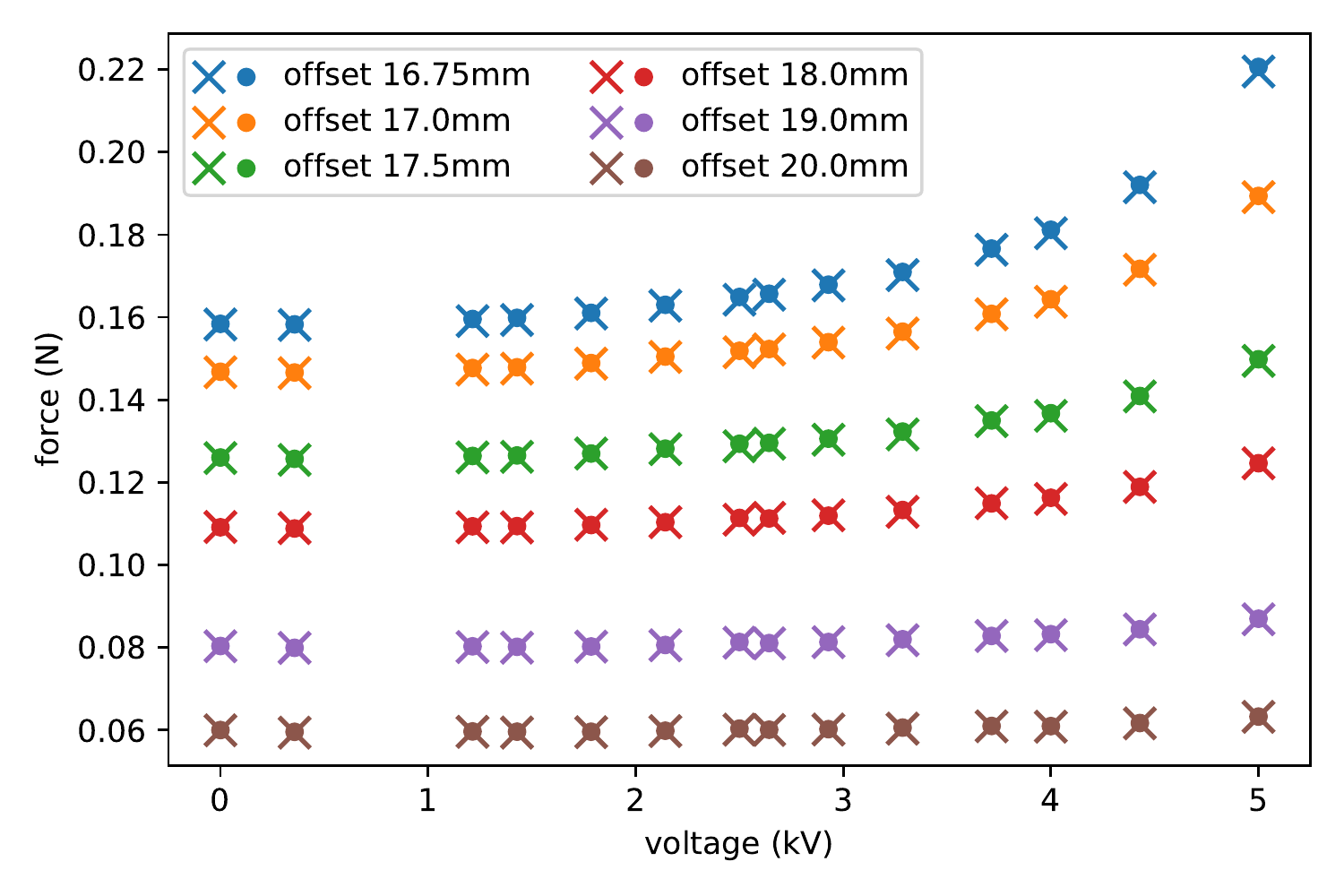}
}
\caption{The steady-state characteristics of DEAP actuator with bias PM-MRE. MRE15 displacement, force (a,b). MRE30 displacement, force (c,d). MRE40 displacement, force (e,f).}
\label{fig:steadyStateMRE-PM}
\end{figure*}

The second stage of experimental work was to show the benefits of using an innovative actuator bias in terms of displacement characteristics in the range of variable supply voltage. To verify the concept of the actuator, individual MRE discs were attached to the DEAP membrane and subjected to the magnetic field from a neodymium permanent magnet, the properties of which are presented in Table \ref{tab:magnetParameters}. The configuration of the actuator is the same as in Figure \ref{fig:actuator}.  The use of the laboratory set presented in Figure \ref{fig:mre-deap-pm-kit} made it possible to obtain the static characteristics of the force between the magnet and the DEAP actuator coupled with individual MRE discs. The laboratory set allows to measure DEAP distance by laser displacement sensor (microEpsilon optoNCDT 1320-10) and force between permanent magnet and magnetorheological elastomer as in the previous measurement. The offset is controlled by the linear module (the gap changes due to the varying voltage of the membrane). The experiment was carried out for different levels of the actuator supply voltage (from \SIrange{0}{5}{\kilo\volt}). Figure \ref{fig:steadyStateMRE-PM} presents the results of the experiments, which were carried out in two independent measurement series. The increase in the distance range for individual voltage values results from the use of the increasing force bias phenomenon presented in Figure \ref{fig:biases}b. When the offset decreases, the force between the permanent magnet and MRE disc increases exponentially, expanding the working range of the DEAP-MRE actuator. The obtained results make it possible to analyze the correlation between the increase in the distance resulting from the Maxwell force and the benefits of the non-linear magnetic bias. Measurement series carried out for various levels of metal flakes powder content in MRE disk prove that it is possible to find offset levels that significantly extend the actuator's operating range. The active offset levels are different for each MRE disk. As shown in Figure \ref{fig:steadyStateMRE-PM}a, for the MRE15, the offset at the level of \SI{20}{\milli\meter} does not cause the actuator to work, while for the MRE40, starting from the voltage of \SI{3}{\kilo\volt}, the operating range of the device is activated (Figure \ref{fig:steadyStateMRE-PM}c).

\begin{figure}
  \centering
   \includegraphics[width=0.6\textwidth]{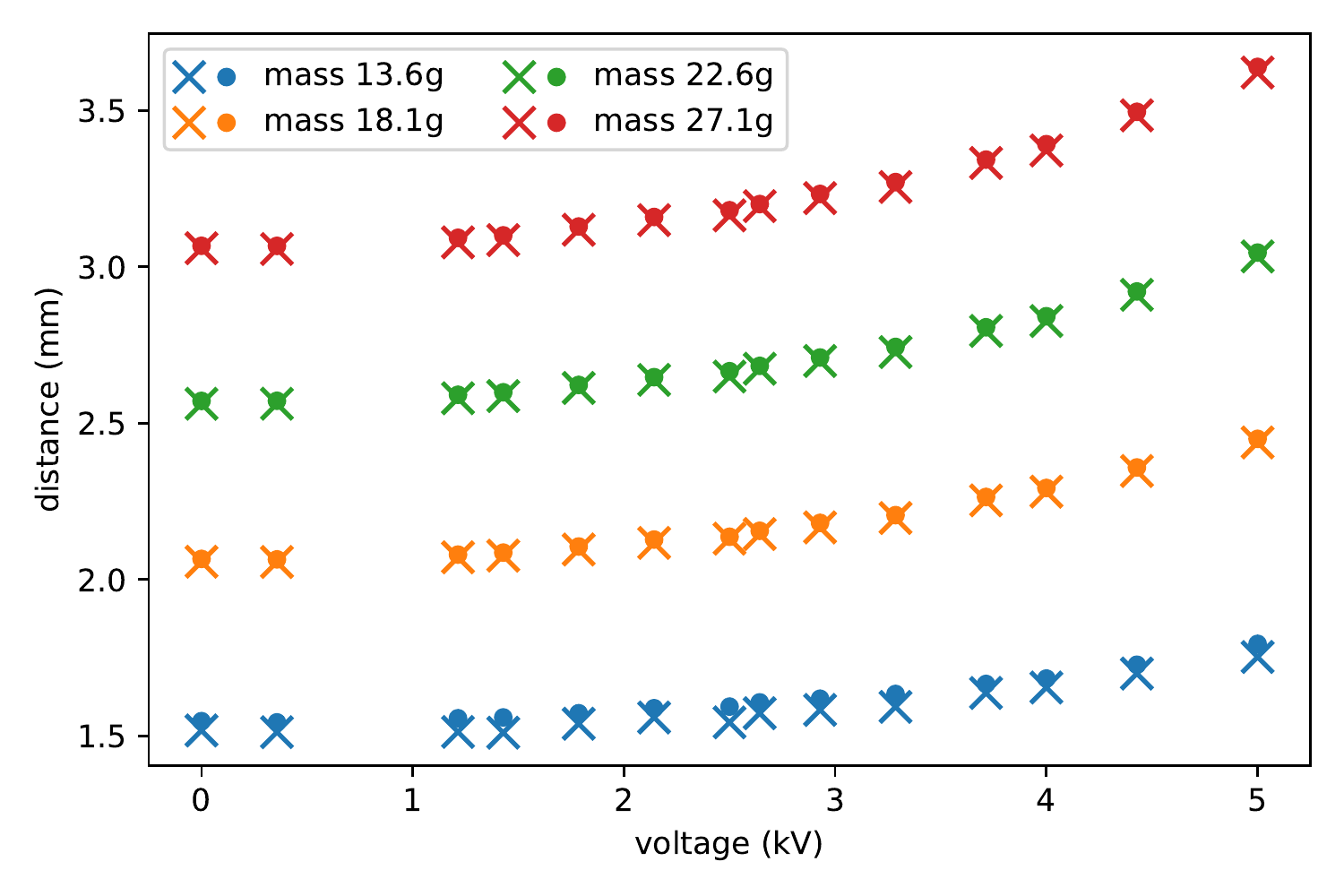}
  \caption{The steady state characteristics between distance and voltage for mass bias.}
  \label{fig:staticMASS}
\end{figure}

In order to show the benefits of MRE-PM bias, the static characteristics for mass bias are also measured for the same membrane and the same range of supply voltages. The results of measurements are visible in Figure \ref{fig:staticMASS}. The experiment also included two measurement series for four exemplary masses that safely used the full operating range of the DEAP actuator membrane. The results of the obtained work ranges were calculated and summarized in Table \ref{tab:workingRange}. The scale of increasing the working range of the DEAP-MRE actuator is very significant in relation to the classic DEAP actuator with mass bias, which allows for more efficient use of the device.

\begin{table}
 \caption{The working range for different type of bias.}
  \centering
  \begin{tabular}{ccc}
    \toprule
    & \multicolumn{2}{c}{Range (mm)}                   \\
    \cmidrule(r){2-3}
    Bias Name & no. try 1 & no. try 2  \\
    \midrule
    PM-MRE15   & 2.154 & 2.272 \\
    PM-MRE30   & 1.819 & 1.979 \\
    PM-MRE40   & 1.308 & 1.321 \\
    Mass 27.1g & 0.564 & 0.572 \\
    Mass 22.6g & 0.471 & 0.471 \\
    Mass 18.1g & 0.382 & 0.385 \\
    Mass 13.6g & 0.242 & 0.251 \\
    \bottomrule
  \end{tabular}
  \label{tab:workingRange}
\end{table}

\subsection{The transients for different initial PM-MRE bias offsets}
\begin{figure*}[!t]
\centering
\subfloat[]{\includegraphics[width=0.45\textwidth]{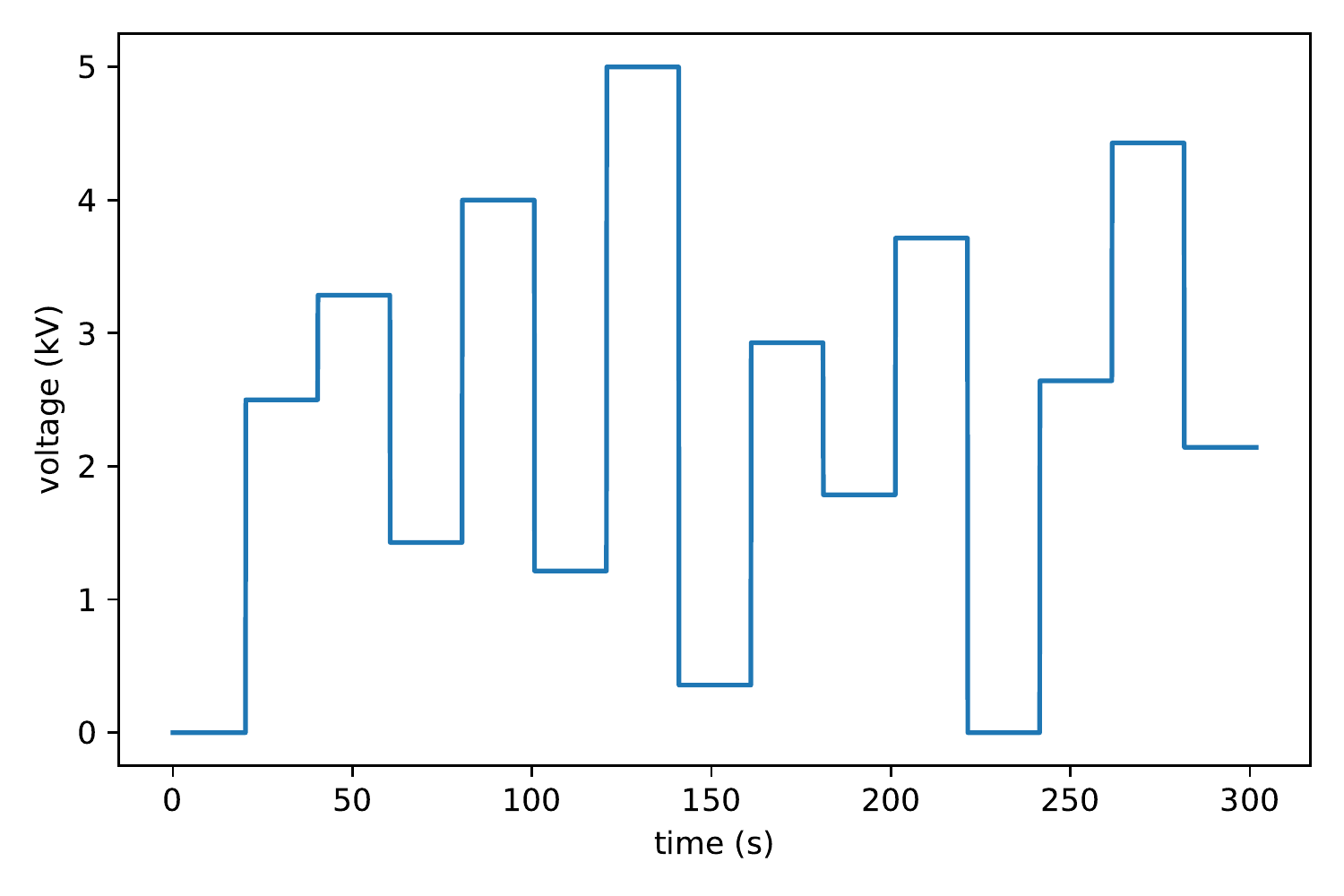}
}
\hfil
\subfloat[]{\includegraphics[width=0.45\textwidth]{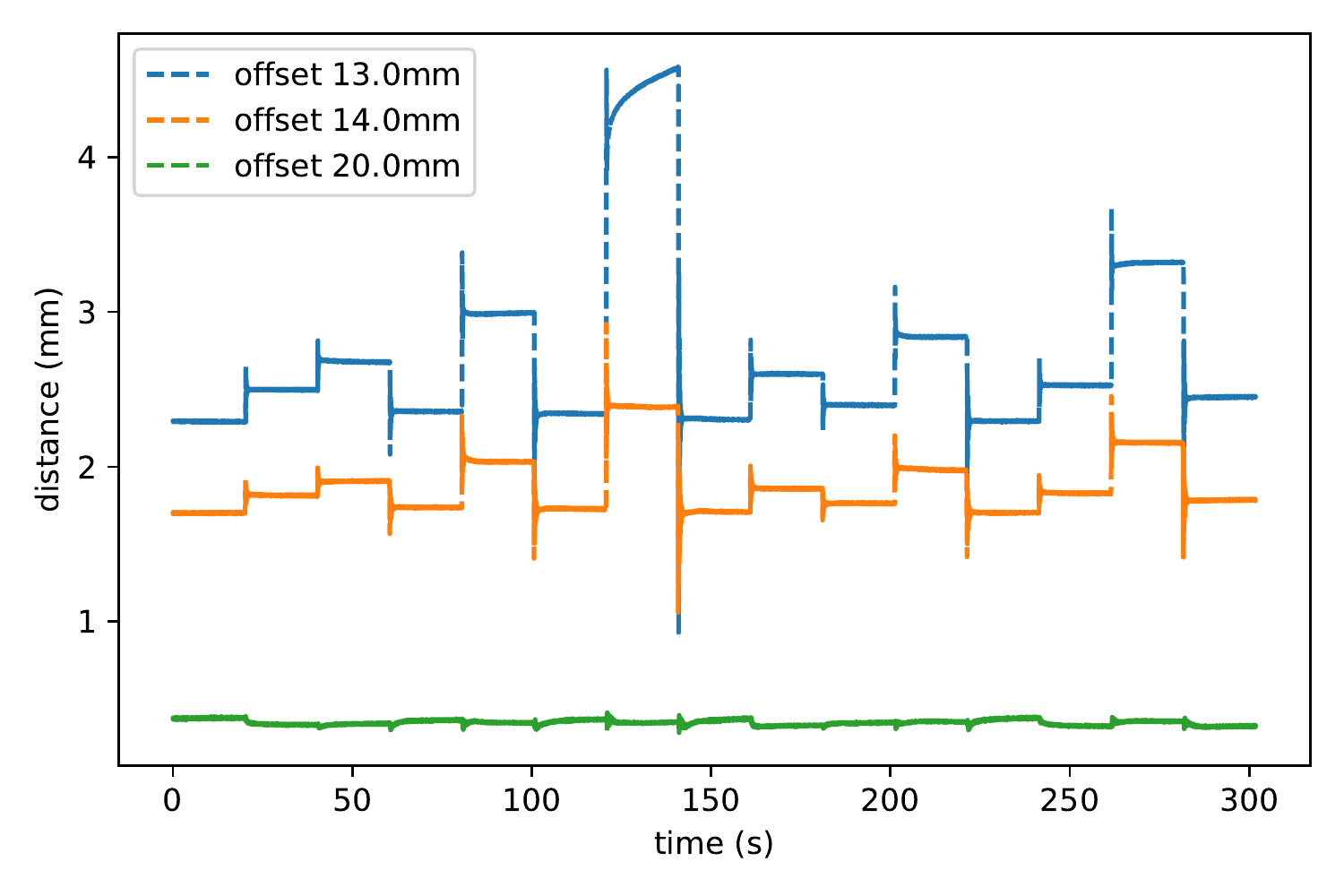}
}
\\
\subfloat[]{\includegraphics[width=0.45\textwidth]{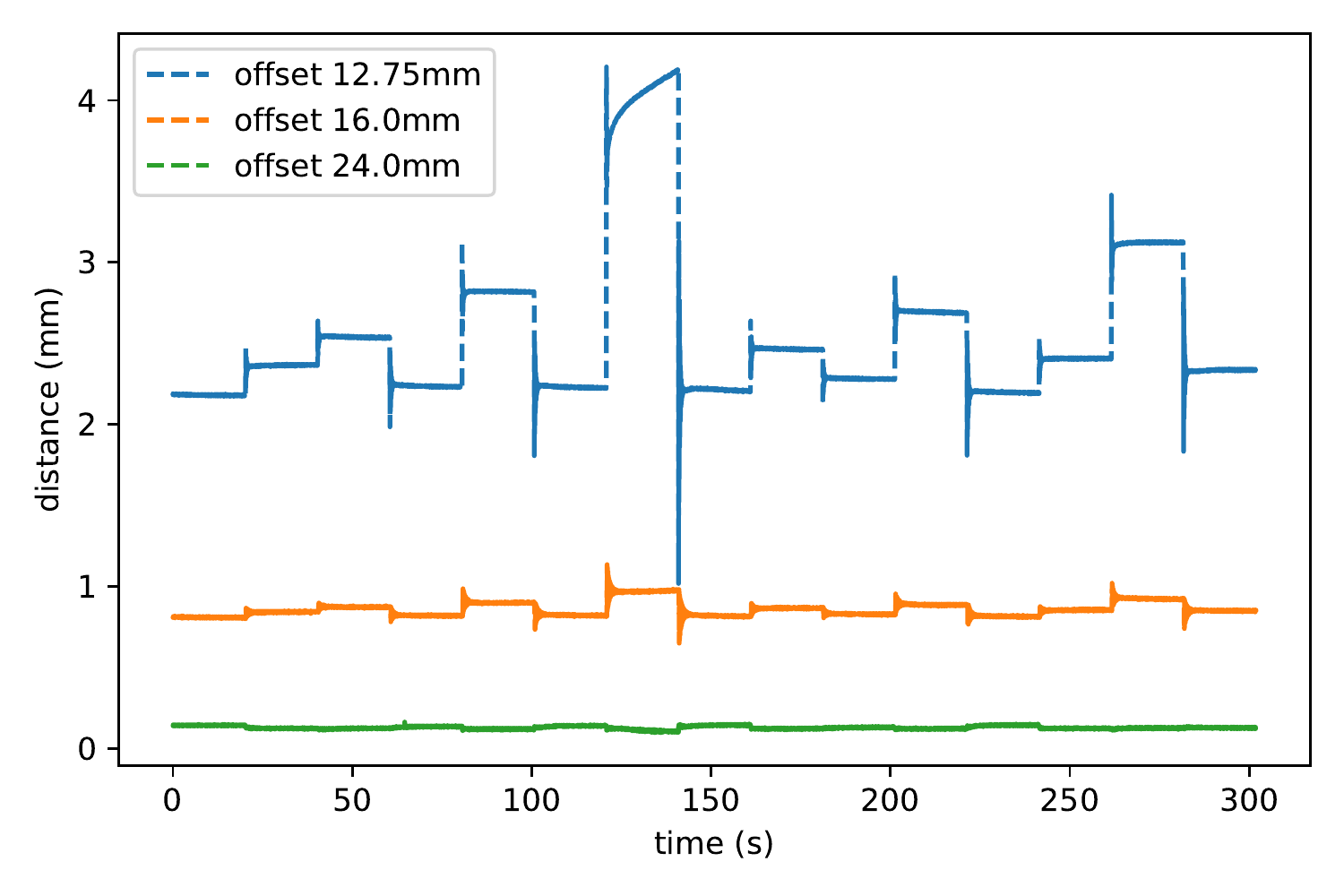}
}
\hfil
\subfloat[]{\includegraphics[width=0.45\textwidth]{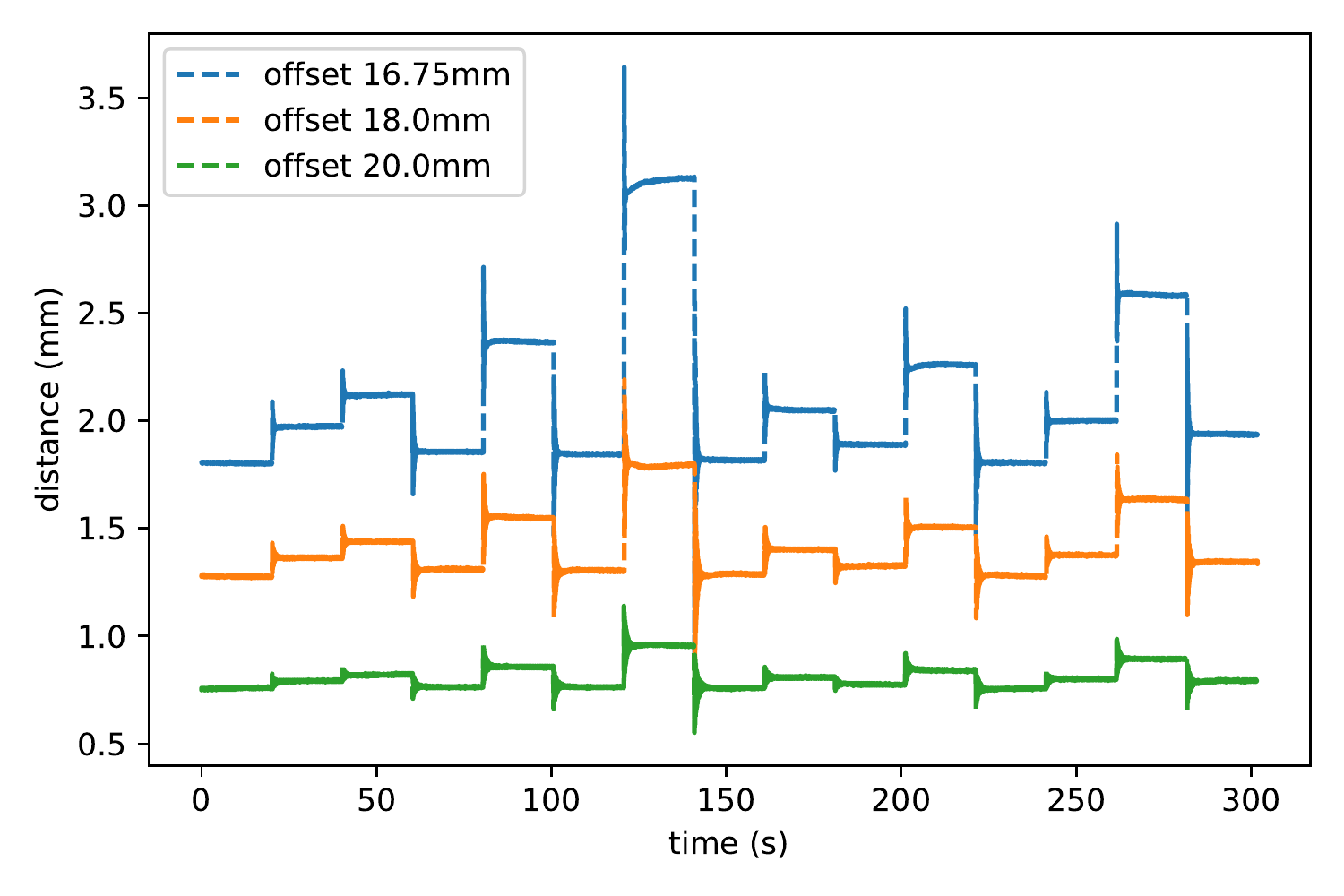}
}
\caption{The transients for different initial offsets for various MRE. The input voltage (a), the output for: MRE15 (b), MRE30 (c), MRE40 (d).}
\label{fig:transientsAll}
\end{figure*}

\begin{figure*}[!t]
\centering
\subfloat[]{\includegraphics[width=0.45\textwidth]{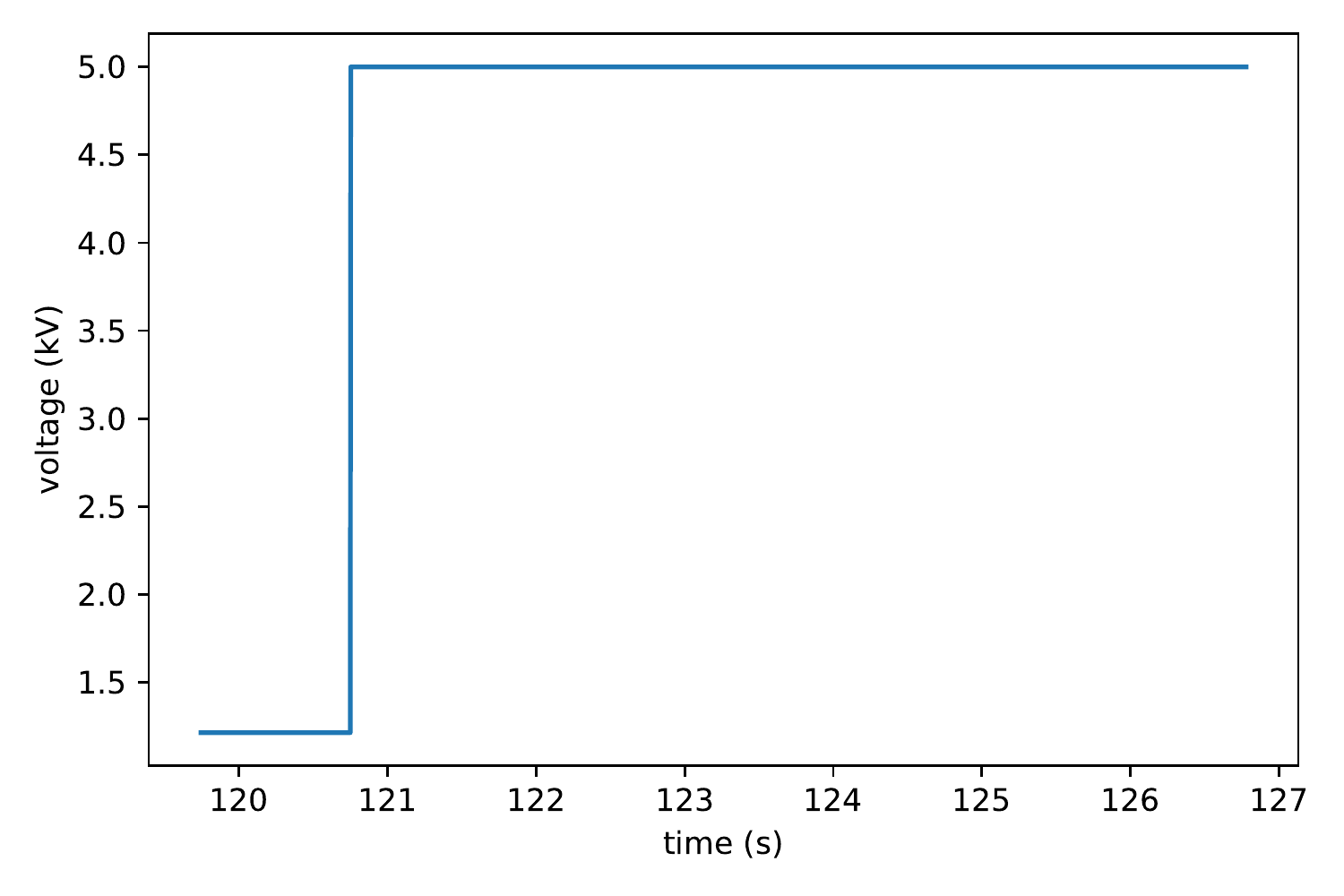}
}
\hfil
\subfloat[]{\includegraphics[width=0.45\textwidth]{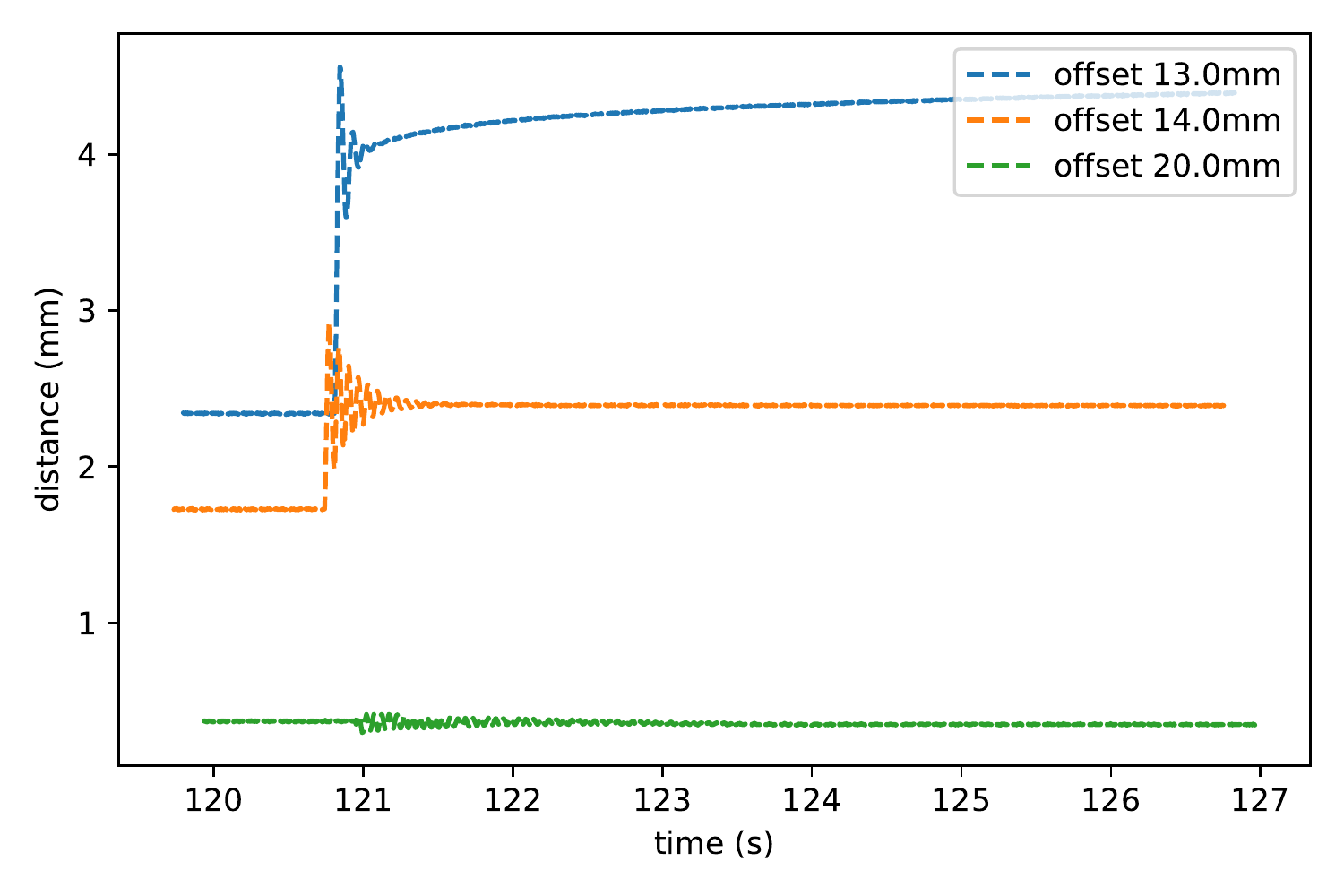}
}
\\
\subfloat[]{\includegraphics[width=0.45\textwidth]{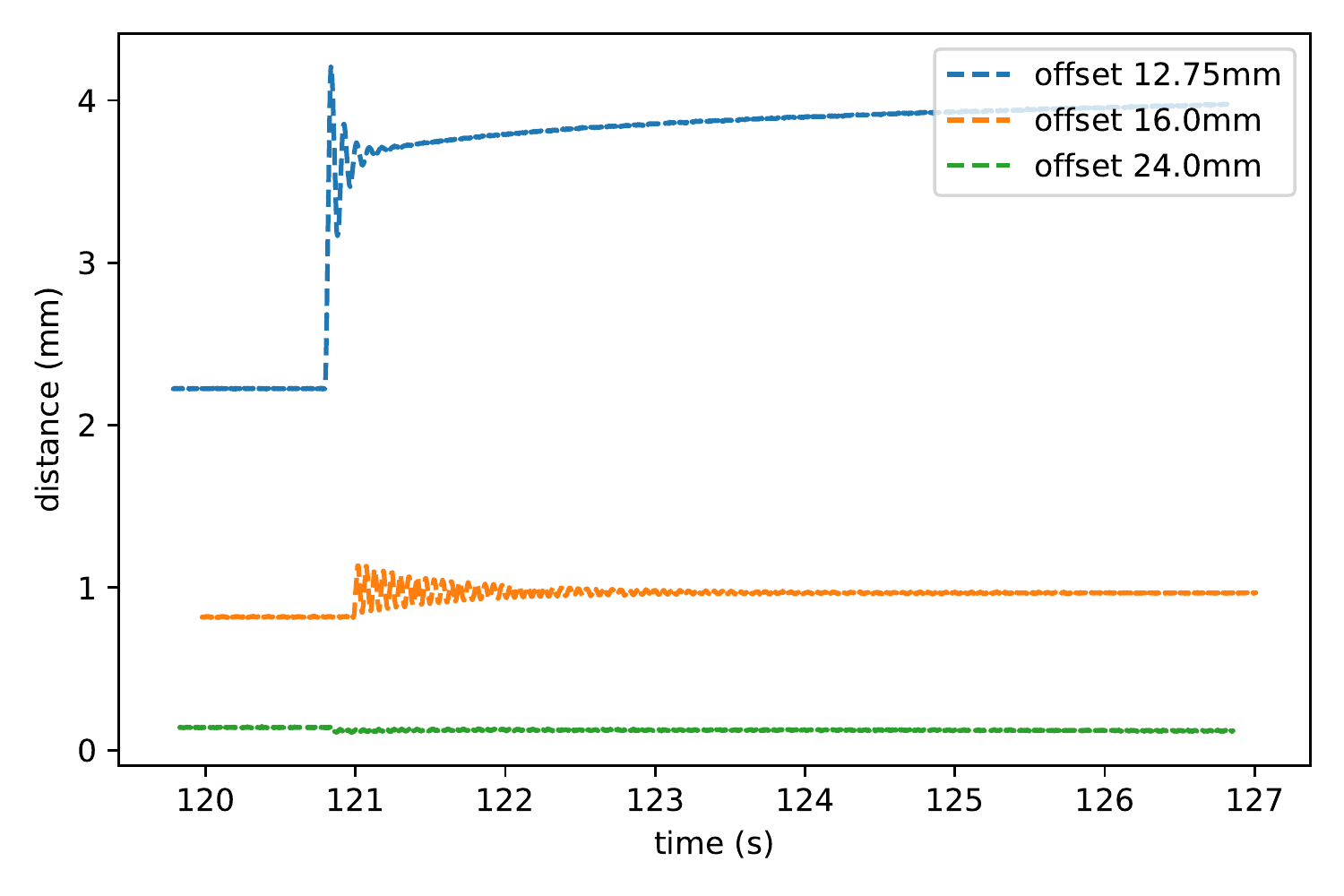}
}
\hfil
\subfloat[]{\includegraphics[width=0.45\textwidth]{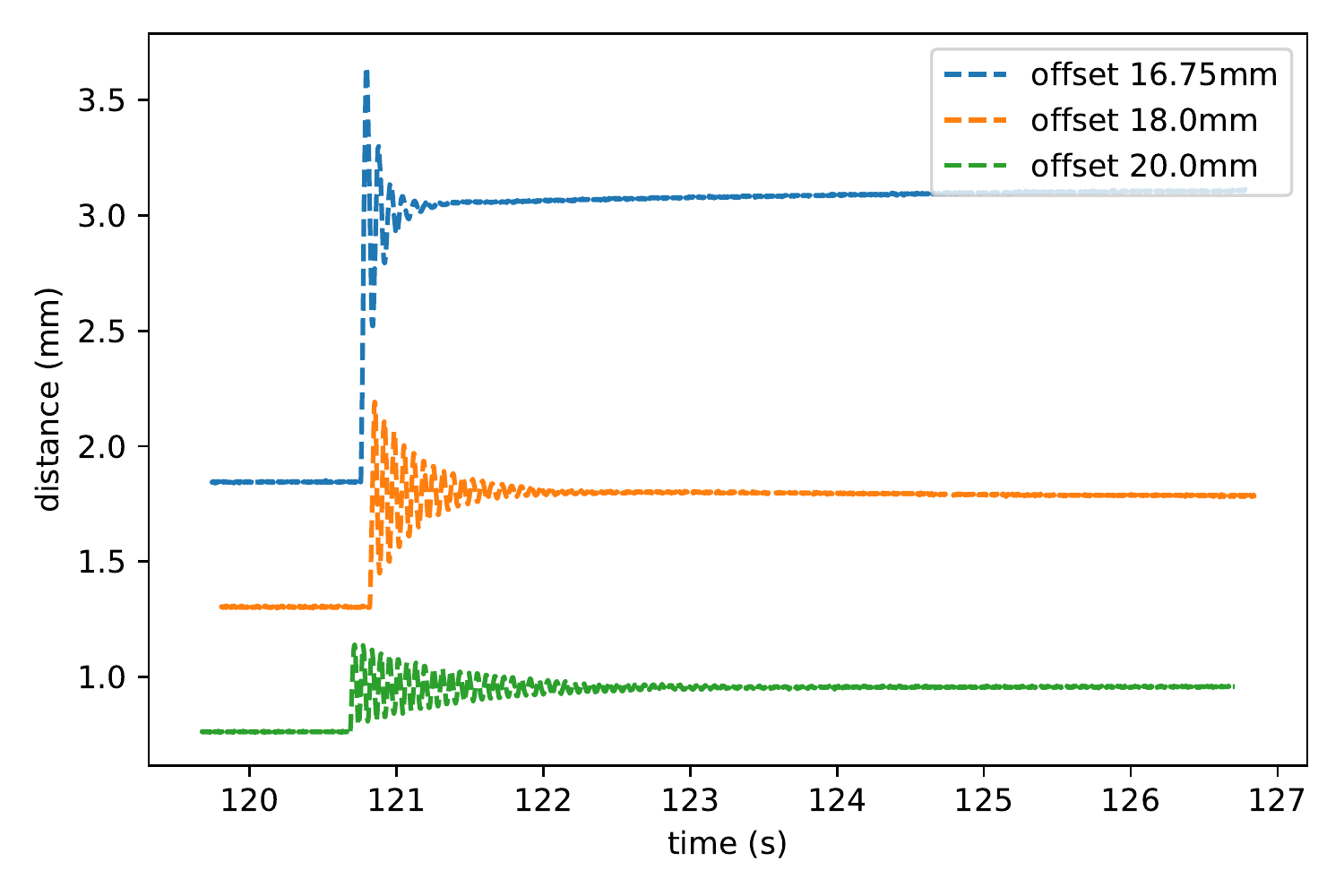}
}
\caption{The zoom of transients for different initial offsets for various MRE. The input voltage (a), the output for: MRE15 (b), MRE30 (c), MRE40 (d).}
\label{fig:transientsZoom}
\end{figure*}

Finally, the operation of the actuator was tested in the field of time-varying voltage step excitation with different amplitudes. The voltage excitation signal is presented in Figure \ref{fig:transientsAll}a, which lasted 300 seconds and included 14-step voltage changes. Figures \ref{fig:transientsAll}b-d show the displacement response of a DEAP actuator coupled with MRE15, MRE30, and MRE40 disks, respectively. Each experiment was carried out for three levels of offsets, which shows the working range of the DEAR-MRE actuators. It is visible that the proper choice of the offset between the permanent magnet and the membrane gives benefits in the higher working range. Furthermore, the general step response of the actuator as shown in Figure \ref{fig:transientsZoom} is consistent with the results obtained in the literature \cite{Sarban:5783345,Rizzello:6867294} because it contains initial oscillations with long relaxation time.

\section{Conclusions}
In this study, we have investigated the novel concept of bias based on the magnetorheological elastomer. The dielectric elastomer actuator works with magnetorheological elastomer with an extended range in comparison to standard mass bias. The difference is up to four times in the best case. The applied bias does not need direct contact due to the magnetic forces. It is also almost independent of gravity because of the small weight of magnetorheological elastomer. It also exploits the non-linear characteristics of the force between MRE and a permanent magnet for various offsets. 

\section*{Acknowledgments}
This research was funded by Ministry of Education and Science, grant number 0311/SIGR/9517

The authors wish to thank the companies HAWA and Mate for providing the metal flake powder used to made MRE disks.

\bibliographystyle{unsrtnat}
\bibliography{references}  






\end{document}